\documentclass[twocolumn,astrosymb,twocolappendix]{aastex631}

\newcommand{\feh}{\mathrm{[Fe/H]}}

\newcommand{\code}[1]{\textbf{\texttt{#1}}}
\newcommand{\artpop}{\code{ArtPop}\xspace}
\newcommand{\sersic}{S\'ersic\xspace}

\usepackage{xspace}
\usepackage{CJKutf8}
\usepackage{bm}
\usepackage{appendix}
\usepackage{amsmath,amssymb}
\usepackage{booktabs}
\usepackage{lipsum}
\usepackage[nomessages]{fp}

\shorttitle{Hedgehog}
\shortauthors{Li et al.}
\graphicspath{{./}{figures/}}

\begin{document}
\begin{CJK*}{UTF8}{gbsn}

\title{Hedgehog: An Isolated Quiescent Dwarf Galaxy at 2.4 Mpc}

\correspondingauthor{Jiaxuan Li}
\author[0000-0001-9592-4190]{Jiaxuan Li (李嘉轩)}
\affiliation{Department of Astrophysical Sciences, 4 Ivy Lane, Princeton University, Princeton, NJ 08544, USA}
\author[0000-0002-5612-3427]{Jenny E. Greene}
\affiliation{Department of Astrophysical Sciences, 4 Ivy Lane, Princeton University, Princeton, NJ 08544, USA}
\author[0000-0002-5382-2898]{Scott G. Carlsten}
\affiliation{Department of Astrophysical Sciences, 4 Ivy Lane, Princeton University, Princeton, NJ 08544, USA}
\author[0000-0002-1841-2252]{Shany Danieli}
\affiliation{Department of Astrophysical Sciences, 4 Ivy Lane, Princeton University, Princeton, NJ 08544, USA}



\begin{abstract}
It is well-known that almost all isolated dwarf galaxies are actively forming stars. We report the discovery of dw1322m2053 (nicknamed Hedgehog), an isolated quiescent dwarf galaxy at a distance of $2.40\pm0.15$ Mpc with a stellar mass of $M_\star \approx 10^{5.8}\, M_\odot$. The distance is measured using surface brightness fluctuations with both Legacy Surveys and deep Magellan/IMACS imaging data. Hedgehog is 1.7~Mpc from the nearest galaxy group, Centaurus A, and has no neighboring galaxies within 1~Mpc, making it one of the most isolated quiescent dwarf galaxies at this stellar mass. It has a red optical color and early-type morphology and shows no UV emission. This indicates that Hedgehog has an old stellar population and no ongoing star formation. Compared with other quiescent dwarfs in the Local Group and Local Volume, Hedgehog appears smaller in size for its luminosity but is consistent with the mass--size relations. Hedgehog might be a backsplash galaxy from the Centaurus A group, but it could also have been quenched in the field by ram pressure stripping in the cosmic web, reionization, or internal processes such as supernova and stellar feedback. Future observations are needed to fully unveil its formation, history, and quenching mechanisms. 
\end{abstract}


\section{Introduction}\label{sec:intro}

Low-mass dwarf galaxies are fragile objects. Their star formation can be halted by internal processes \citep[e.g., stellar feedback;][]{Hopkins2012,ElBadry2018} and external processes \citep[e.g., tidal and ram pressure stripping;][]{GunnGott1972,Simpson2018}, and reionization \citep[e.g.,][]{Bullock2000,Ricotti2005,Applebaum2021}. Consequently, dwarf galaxies serve as sensitive probes for understanding both baryonic physics and environmental effects.

All satellite dwarf galaxies of the Milky Way (MW), except for the Large and Small Magellanic Clouds, are quiescent \citep[e.g.,][]{McConnachie2012,Karachentsev2013,Wetzel2015}. In contrast, quiescent dwarfs are found to be exceptionally rare ($<0.06\%$) among the isolated dwarfs with $M_\star = 10^{7-9}\, M_\odot$ \citep{Geha2012}. Despite their rarity, several isolated quiescent dwarf galaxies have been discovered. Some are located on the outskirts of the Local Group (e.g., Tucana, \citealt{Saviane1996}; Cetus, \citealt{McConnachie2006}; And XVIII, \citealt{McConnachie2008}; Eri II, \citealt{Crnojevic2016}) and M81 (e.g., Blobby, \citealt{Casey2023}). Others are found far from any massive galaxy (e.g., KKR 25, \citealt{Makarov2012}; KKs 3, \citealt{Karachentsev2015_KKs3}; COSMOS-dw1, \citealt{Xi2018,Polzin2021}; UGC~5205, \citealt{Kado-Fong2024})\footnote{We note that PEARLSDG \citep{Carleton2024}, which was initially thought to be an isolated quiescent dwarf galaxy, is recently confirmed to be a background galaxy and is associated with a galaxy group \citep{Carleton2024b}}. While the majority of these dwarfs are confirmed to be in the field by measuring distances using the tip of the red giant branch (TRGB), the surface brightness fluctuation \citep[SBF;][]{Tonry1988,Carlsten2019,Greco2021} technique has recently become an efficient new method for discovering and confirming these galaxies, including COSMOS-dw1 \citep{Polzin2021} and Blobby \citep{Casey2023}.

The mechanisms by which these isolated dwarfs are quenched remain unclear. One possible explanation is that these dwarfs might have already entered the virial radius of a bigger group, experienced one pericenter passage, and been ejected into the field. Consequently, they could be quenched due to close interactions with the host halo \citep{Teyssier2012}. Simulations suggest that these so-called ``backsplash'' dwarf galaxies can be found out to $\sim 2.5$ times the virial radius \citep{Wetzel2014,More2015,Buck2019,Applebaum2021,Benavides2021}. On the other hand, isolated dwarf galaxies could be quenched via ram pressure when they move through the cosmic web \citep{Benitez-Llambay2013}. Stellar feedback might also temporarily halt star formation \citep{ElBadry2018}. Ultra-faint dwarfs in the field can be quenched by reionization alone \citep[e.g., Tucana B;][]{Sand2022}. 

In this Letter, we report the discovery of dw1322m2053 (nicknamed Hedgehog\footnote{We name it ``Hedgehog'' because hedgehogs are small and solitary animals.}), an isolated quiescent dwarf galaxy at a distance of $D\approx 2.4$~Mpc. With no neighboring galaxies within 1~Mpc and located 1.7~Mpc from the nearest massive galaxy group ($4-5$ times the virial radius of the group), Hedgehog is one of the most isolated quiescent dwarfs found to date. We present the discovery of this galaxy and our follow-up observations in Section \ref{sec:discovery}. In Section \ref{sec:distance}, we measure the SBF distance to this dwarf galaxy using data from the DESI Legacy Imaging Surveys Data Release 10 \citep[hereafter Legacy Surveys or LS DR10;][]{Dey2019} and the deep high-resolution Magellan Inamori Magellan Areal Camera and Spectrograph (IMACS) imaging data. We show the 3D environment of Hedgehog in Section \ref{sec:environment}, then present the physical properties and characterize the stellar population of Hedgehog in Section \ref{sec:sp}. We discuss possible quenching mechanisms in Section \ref{sec:discuss}.

We adopt a flat $\Lambda$CDM cosmology with $\Omega_{\rm m}= 0.3$ and $H_0 = 70\ $km s$^{-1}$ Mpc$^{-1}$. The virial radius ($R_{\rm vir}$) of a dark matter halo is defined as the radius within which the average density is $\Delta = 200$ times the critical density, and the virial mass $M_{\rm vir}$ is the enclosed mass within $R_{\rm vir}$. All photometry presented in this work is in the AB system \citep{Oke1983}. We apply corrections for the MW dust extinction using the dust map in \citet{SFD1998} recalibrated by \citet{Schlafly2011}. The solar absolute magnitudes used in this work are taken from \citet{Willmer2018}.

\begin{figure*}[t]
    \centering
    \includegraphics[width=0.85\linewidth]{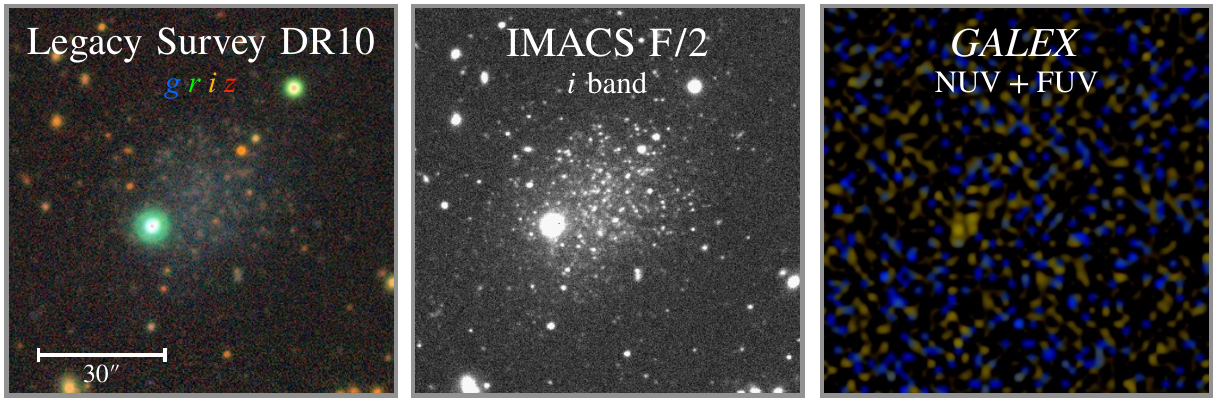}
    \caption{Cutout images of Hedgehog from the Legacy Surveys DR10 (left), the deep and high-resolution Magellan/IMACS $i$-band data (middle), and the Galaxy Evolution Explorer (\textit{GALEX}) near-UV and far-UV data (right). Hedgehog shows significant surface brightness fluctuation signals in the Legacy Surveys data and is resolved into individual stars in the IMACS data. There is also no significant UV emission associated with Hedgehog.}
    \label{fig:image}
\end{figure*}

\section{Discovery, Observations, and Photometry}\label{sec:discovery}
Hedgehog was discovered serendipitously in the Legacy Surveys DR10 data when we searched for potential dwarf galaxies associated with NGC~5068 at $D = 5.15$~Mpc \citep{Karachentsev2018}. Located $\sim1\deg$ away from NGC~5068, Hedgehog was initially considered a satellite candidate of NGC~5068. As shown in the left panel of Figure \ref{fig:image}, Hedgehog exhibits significant SBF in Legacy Surveys DR10 data, strongly suggesting that it might be a foreground dwarf galaxy. It also has a red color and an early-type morphology without any star-forming regions, making it a possible candidate for an isolated quenched dwarf galaxy.

\subsection{Follow-up observations}

To determine the distance and the properties of Hedgehog, we conducted follow-up observations with IMACS \citep[][]{Dressler2011} on the 6.5 m Magellan Baade Telescope on 2024 April 1. Using the $f/2$ camera, which offers a field of view of $27.5\arcmin$ in diameter and a pixel scale of $0.2\arcsec\ \mathrm{pixel}^{-1}$, we obtained deep imaging data for Hedgehog in the Sloan $i$-band filter. We took four exposures of 5 minutes each and dithered between exposures to bridge the chip gaps. The seeing was about $0.5-0.6\arcsec$ during the observation. We reduced the data using a custom pipeline similar to that used in \citet{CarlstenELVES2022}. It includes bias subtraction and flat-fielding correction using the twilight flat frames. The astrometric solution was obtained using \textsc{astrometry.net} \citep{astrometry_net}, and the local sky background was subtracted using \textsc{SExtractor} \citep{Bertin1996} with a mesh size of 51.2\arcsec. We calibrated our photometry against the DECam Local Volume Exploration Survey \citep[DELVE;][]{DELVE2021} DR2 photometric catalog \citep{DELVE2022}, such that all photometry in this work is in the DECam filter system. In the end, we fine-tuned the astrometric solution using \textsc{scamp} \citep{scamp} and coadded all exposures using \textsc{swarp} \citep{swarp}. A spatially varying point-spread function (PSF) model was constructed using \code{PSFEx} \citep{psfex} and evaluated at the location of Hedgehog. The final coadded $i$-band image has a PSF FWHM of $0.6\arcsec$, indicating that the PSF is marginally undersampled. 

We show the coadded IMACS image in the middle panel of Figure \ref{fig:image}. Hedgehog is resolved into individual stars thanks to the excellent seeing and depth. Unfortunately, we do not have follow-up data in bluer bands to construct a color-magnitude diagram (CMD) and determine the TRGB distance. As a result, we seek to measure distance using the SBF technique. 

SBF measurements are sensitive to the noise properties of the image. As demonstrated by \citet{Mei2005}, different interpolation kernels used during resampling in the coaddition process would significantly change the noise power spectrum and consequently affect the distance measurement. A commonly used \code{Lanczos3} kernel introduces noise correlation, causing the noise power spectrum to drop at a spatial frequency of $k \gtrsim 0.35\ \rm{pixel}^{-1}$ \citep{Cantiello2005,Carlsten2019}. This makes fitting for the noise level challenging, especially when the PSF is sharp and undersampled. To mitigate this issue, we use the nearest-neighbor interpolation kernel, which preserves the noise property of the original images. Since Hedgehog is well covered by all the exposures, we do not find any aliasing or empty pixels in the coadded image around Hedgehog.

\begin{table}
\centering
\caption{Properties of Hedgehog. \label{tab:property}}
\begin{tabular}{lr}
\toprule
Parameter & Value \\
\midrule
$\alpha_0$ (J2000) & 13:22:46.88 \\
$\delta_0$ (J2000) & $-$20:53:55.94 \\
$m_g$ (mag) & $17.35 \pm 0.06$ \\
$g-r$ (mag) & $0.49 \pm 0.05$ \\
$g-i$ (mag) & $0.62 \pm 0.05$ \\
$r_{\rm eff}$ (arcsec) & 15.1 $\pm$ 0.8 \\
$n_{\rm S\acute{e}rsic}$ & $0.56 \pm 0.03$\\
$\mu_{0,g}$ (mag arcsec$^{-2}$) & $24.80 \pm 0.15$ \\
Ellipticity & $0.22 \pm 0.03$ \\
$\mathrm{S/N}_{\rm NUV}$ & 1.8 \\
$m_{\rm NUV}$ (mag) & $>17.5$ \\ 
$\mathrm{S/N}_{\rm FUV}$ & 0.7 \\
$m_{\rm FUV}$ (mag) & $>19.00$ \\ 
\midrule
Distance (Mpc) & $2.40 \pm 0.15$ \\
$M_g$ (mag) & $-9.56 \pm 0.14$ \\
$M_V$ (mag) & $-9.84 \pm 0.16$ \\
$\log(M_\star/M_\odot)$ & $5.8 \pm 0.2$ \\
$r_{\rm eff}$ (pc) & 176 $\pm$ 14 \\
$\log\,(\rm SFR_{NUV}/M_\odot\, \text{yr}^{-1})$ & $<-3.7$ \\ 
$\log\,(\rm SFR_{FUV}/M_\odot\, \text{yr}^{-1})$ & $<-4.5$ \\ 
$\log\,(M_{\rm HI}/M_\odot)$ & $<6.0$ \\
\bottomrule
\end{tabular}
\end{table}

\subsection{Optical Photometry and Structural Properties}\label{sec:sersic}
We measure the structural properties of Hedgehog by fitting a single \sersic model to the Legacy Surveys DR10 data in the $gri$ bands. This part of the Legacy Surveys data is primarily from the DELVE survey \citep{DELVE2021,DELVE2022}. It is shallower and has much worse seeing compared with our IMACS data\footnote{The seeing values are $1.3\arcsec, 2.3\arcsec, 1.3\arcsec$ in the $gri$ bands, and the 5$\sigma$ PSF detection depths are 24.9, 24.3, and 23.9 mag, respectively.}. Similar to \citet{ELVES-I}, we directly take the coadded images and the corresponding PSF models from the Legacy Surveys DR10 for \sersic fits using \textsc{imfit} \citep{imfit}. 

We start with fitting a \sersic model to the $g$-band image because it is deeper and has the fewest SBF signals \citep{Carlsten2019,Greco2021}. The free parameters include the central right ascension ($\alpha_0$) and declination ($\delta_0$), total $g$-band magnitude ($m_g$), effective radius ($r_{\rm eff}$, measured along the semi-major axis), \sersic index ($n_{\rm S\acute{e}rsic}$), and ellipticity (defined as $1-b/a$). Subsequently, we fit the $r$- and $i$-band images using the best-fit \sersic model from the $g$ band, fixing the structural parameters and allowing only the total magnitude to vary. To estimate potential measurement uncertainties and biases in photometry and structural parameters, we inject mock \sersic{} galaxies into the Legacy Surveys DR10 coadded images and fit them using the same procedure described above (see Appendix \ref{ap:bias} for details). We correct for measurement biases and list the photometry and structural parameters in Table \ref{tab:property}.

\subsection{Galaxy Evolution Explore}\label{sec:galex}
We search for possible UV emission from Hedgehog using data from the Galaxy Evolution Explorer \citep[\textit{GALEX};][]{Martin2005} survey, following the method described in \citet{Greco2018b}, \citet{Karunakaran2021}, and \citet{CarlstenELVES2022}. We first query the available images that overlap with Hedgehog on the Mikulski
Archive for Space Telescopes (MAST) \textit{GALEX} tile retrieval service\footnote{\url{https://galex.stsci.edu/GR6/?page=tilelist&survey=ais&showall=Y}}. Tile \code{AIS\_234\_sg74}, which has the longest exposure time ($t_{\rm NUV} = 207.05$ s,  $t_{\rm FUV} = 202.05$ s) among the available tiles, is chosen for photometry. We download the intensity image ($I$) and the high-resolution relative response maps ($R$) from MAST and calculate the variance image $V = I / R$.

Using \textsc{photutils} \citep{photutils}, we perform circular aperture photometry for Hedgehog with a radius twice the effective radius from the \sersic fits in \S\ref{sec:sersic}. This aperture size includes most of the flux from the galaxy but not too much background noise. Since the intensity image is not background-subtracted, we estimate the background value and its standard deviation. After masking out all bright sources using \code{sep} \citep{Barbary2016}, we randomly place 100 apertures within 6 arcmin from the target galaxy. We calculate the median of the values in these apertures and subtract it from the intensity image. The standard deviation, $\sigma_{\rm sky}$, represents the uncertainty from the sky background. We then do aperture photometry at the location of Hedgehog to measure the flux $F = \sum_i I_i$, where $i$ ranges over all pixels in the aperture. The bright star near Hedgehog is masked. The uncertainty of the flux is then $\sigma_F = (\sum_i V_i + \sigma_{\rm sky}^2)^{1/2}$, and the signal-to-noise ratio (S/N) is defined as $\mathrm{S/N} = F / \sigma_F$. We convert the measured flux to AB magnitude taking the zero-points from \citet{Morrissey2007}, and correct for Galactic extinction using $R_{\rm NUV}=8.2$ and $R_{\rm FUV}=8.24$ \citep{Wyder2007}. 

The \emph{GALEX} photometry results are listed in Table \ref{tab:property}. We do not find significant emission in either near-UV (NUV) or far-UV (FUV), with $\mathrm{S/N} < 2$ for both bands. Therefore, we report the $2\sigma$ lower limits of the magnitudes in Table \ref{tab:property}. We note that the \textit{GALEX} data used here are relatively shallow compared with the data used in other similar studies \citep[e.g.,][]{Sand2022,Karunakaran2022,CarlstenELVES2022}.

\begin{figure*}
    \centering
    \includegraphics[width=0.9\linewidth]{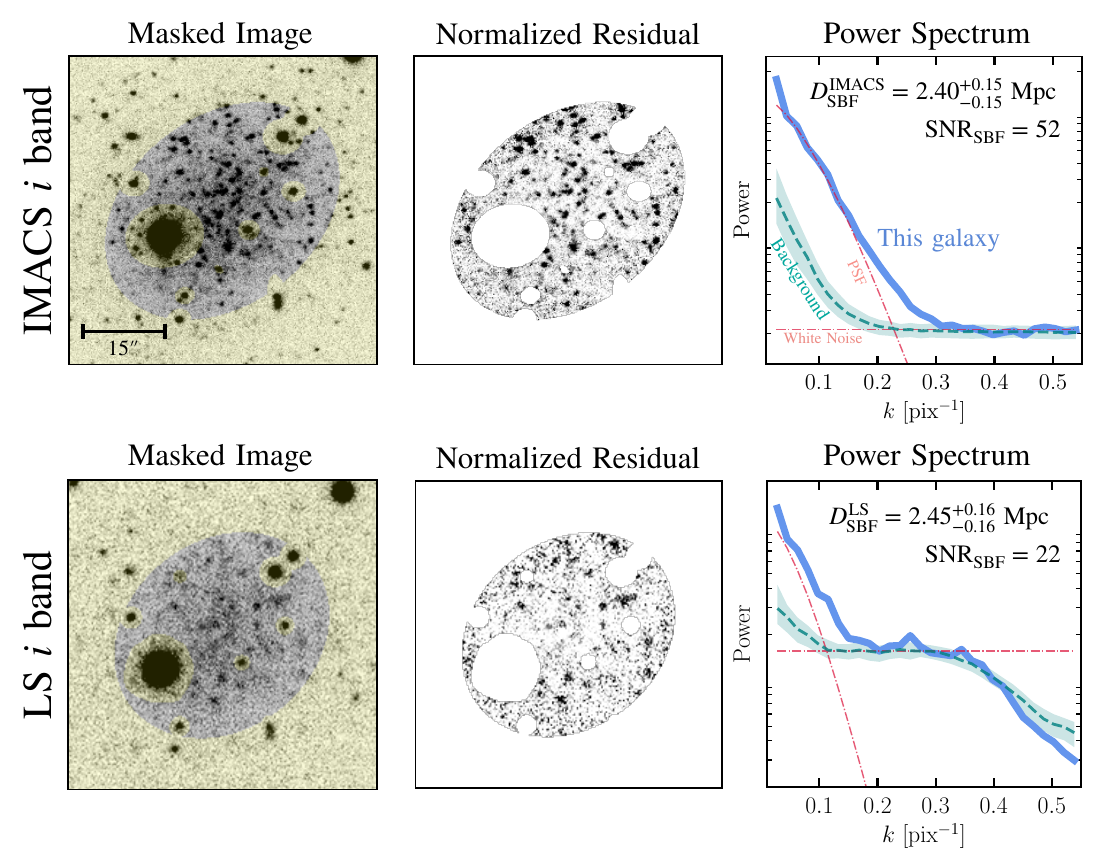}
    \vspace{-1em}
    \caption{The SBF measurements for Hedgehog using the IMACS $i$-band data (top panels) and the Legacy Surveys DR10 $i$-band data (bottom panels). The left panels show the original image with the mask overlaid in yellow. We fit a \sersic model to the masked image, subtract the model from the image, and divide the residual by the square root of the model. This normalized residual image is shown in the middle panels. We then calculate the azimuthally averaged power spectrum of the normalized residual, as shown in the right panels. The power spectrum is fit with a combination of the PSF power spectrum and white noise. To account for the contribution from unmasked sources, we measure SBF signals on randomly selected blank fields, whose power spectra are shown as the teal dashed line. After subtracting the background contribution, we calculate the median SBF signal and its standard deviation and convert it to distance using the SBF--color relation in \citet{Carlsten2019}. The SBF distances from IMACS and Legacy Surveys DR10 are $D_{\rm SBF}^{\rm IMACS} = 2.40 \pm 0.15\ \rm Mpc$, $D_{\rm SBF}^{\rm LS} = 2.45 \pm 0.16\ \rm Mpc$. These two independently measured distances are fully consistent.}
    \label{fig:sbf_figure}
\end{figure*}

\section{Distance}\label{sec:distance}
For Hedgehog, distance is the most crucial piece of information needed to determine its physical properties and environment. In this section, we measure the SBF distance to Hedgehog \textit{independently} using both IMACS and Legacy Surveys DR10 $i$-band data. Here we briefly summarize the SBF measurement and refer interested readers to \citet{Carlsten2019,ELVES-I}, \citet{Greco2021}, \citet{Kim2021} and references therein for more detailed descriptions of the SBF technique in the dwarf galaxy regime. 

SBF measures the pixel-to-pixel variation in the image due to the Poisson fluctuations in the number of bright stars per resolution element. As distance increases, the number of stars per pixel grows, and the variance decreases. Thus, the SBF signal depends on distance and the stellar population of the galaxy. In addition, the fluctuation from the background galaxies and other unresolved sources contributes to the measured fluctuation and needs to be masked or corrected statistically.

In this work, we measure the SBF distance in the $i$ band since the SBF signal is stronger and the seeing is typically better. In practice, we first fit a smooth \sersic model to the target galaxy (similar to \S\ref{sec:sersic}), subtract it from the image, and normalize the residual image by dividing by the square root of the \sersic model. To mask out globular clusters, background galaxies, and foreground stars, we use a certain absolute magnitude threshold $M_i$ \citep{Carlsten2019}. Sources brighter than this threshold are masked. In this work, we set $M_i^{\rm IMACS} = -4.0$~mag for IMACS data and $M_i^{\rm LS} = -4.2$~mag for Legacy Surveys data assuming a distance of 2.4 Mpc. This ensures that globular clusters ($M_i \sim -8$~mag) are masked, while most of the asymptotic giant branch (AGB) stars and the brightest red giant branch stars ($M_i\sim -4$~mag) are kept from masking. In a low-mass system like Hedgehog, the number of AGB stars brighter than $M_i=-4.0$~mag is very small, and they contribute $<1\%$ to the total SBF signal. It is thus safe to mask out a few very bright AGB stars if there are any.

Because the distance of Hedgehog is unknown, we iteratively adjust the assumed distance and the absolute magnitude threshold until it converges. The absolute magnitude thresholds are different for IMACS and Legacy Surveys because of their different resolution and depth. Using a fainter threshold for the Legacy Surveys data will mask too many regions contributing to the SBF signal. To maximize the S/N, we also mask out regions where the surface brightness of the galaxy falls below 0.25 times the central surface brightness. The yellow color in the left panels of Figure \ref{fig:sbf_figure} highlights the mask. The IMACS and Legacy Surveys masks agree with each other quite well. The masked sources include a bright foreground star and likely several Galactic M dwarf stars because of their very red colors (Figure \ref{fig:image}). The middle panel shows the masked normalized residual maps.

We compute the azimuthally averaged power spectrum of the masked normalized residual image, shown as the blue solid line in the right panel of Figure \ref{fig:sbf_figure}. We then fit it with a combination of the power spectrum of the PSF and a constant white-noise floor (red dashed lines in Figure \ref{fig:sbf_figure}). The desired SBF signal is the coefficient of the PSF component. For dwarf galaxies with low surface brightness, the unmasked contaminating sources would contribute significantly to the measured SBF signal. We further estimate the contribution from the residual contaminating sources by measuring the SBF signals from randomly selected blank fields around the target galaxy. For blank fields, we construct the normalized residual image using the \sersic model from the target galaxy and use the same masking threshold to generate masks. The signals from the blank fields are subtracted from the measured SBF signal. The teal-shaded region in Figure \ref{fig:sbf_figure} demonstrates the power spectrum of the background fields.

To characterize the uncertainty of the SBF measurement, we adopt a Monte Carlo approach by randomly sampling the location of the background fields and the spatial frequency range used for the power spectrum fitting \citep{Cohen2018}\footnote{For IMACS, we also tried sampling the absolute magnitude threshold for generating the mask $M_i \sim \mathcal{U}(-3.9, -4.1)$. It reduces the S/N of the SBF signal by a factor of 1.5 compared to fixing $M_i=-4.0$. However, it only increases the distance uncertainty by 0.02 Mpc.}. The lower spatial frequency follows $k_1 \sim \mathcal{U}(0.03, 0.08)\ \rm{pixel}^{-1}$. The upper spatial frequency follows $k_2 \sim \mathcal{U}(0.35, 0.45)\ \rm{pixel}^{-1}$ for IMACS but follows $k_2 \sim \mathcal{U}(0.3, 0.4)\ \rm{pixel}^{-1}$ for Legacy Surveys because the correlated noise in the Legacy Surveys image becomes prominent around $k\gtrsim 0.35\ \rm{pixel}^{-1}$. We do 100 Monte Carlo runs and take the median value of the background-corrected SBF signals to be the measured SBF signal and their standard deviation to be the uncertainty. The SBF S/N is defined as the SBF signal divided by the uncertainty. We then use the SBF--color relation to convert the measured SBF signal to distance. The distance uncertainty is derived using Monte Carlo simulations, taking into account the uncertainties in the measured galaxy color, the SBF signal, and the scatter in the SBF--color relation \citep{Carlsten2019}. The difference between the Magellan/IMACS and Blanco/DECam $i$-band filters (typically $\lesssim 0.1$~mag difference in the SBF signal; \citealt{CarlstenELVES2022}) is neglected because it is less dominant than the scatter of the SBF--color relation and the measurement uncertainty.


We take the SBF--color relation in \citet{Carlsten2019} as our fiducial model to convert the SBF signal to distance. We note that the intercept and slope in their Equation 4 are highly correlated; thus, we sample their joint distribution\footnote{The joint distribution of the intercept and slope of the SBF--color relation in \citet{Carlsten2019} can be approximated by a multivariant Gaussian distribution with a mean of $\mathbf{\mu} = (-3.18,\ 2.15)$ and a covariance matrix of $\Sigma = \begin{pmatrix} 0.0461 & -0.0839 \\ -0.0839 & 0.1671 \end{pmatrix}$.} to derive the distance uncertainty. Using the IMACS $i$-band data, the measured SBF distance to Hedgehog is $$D_{\rm SBF}^{\rm IMACS} = 2.40 \pm 0.15\ \rm Mpc,$$ with an S/N of $\rm S/N \approx 52$. When using the Legacy Surveys DR10 $i$-band data, the measured SBF distance is $$D_{\rm SBF}^{\rm LS} = 2.45 \pm 0.16\ \rm Mpc,$$ with an S/N of $\rm S/N \approx 22$. The distance measurements from two independent datasets perfectly agree with each other. Because the SBF distances from IMACS and Legacy Surveys are very consistent, we take the IMACS distance as the fiducial value hereafter. This distance corresponds to a distance modulus of $m-M = 26.90 \pm 0.13$~mag.

We also explored other calibrations for the SBF--color relation using the IMACS data. Using the calibration in \citet{Kim2021} with $M_{g,\rm thres}=-4.0$~mag, we get $D_{\rm SBF}=2.40 \pm 0.23$~Mpc, assuming the intercept and slope are independent random variables. Using the relation in \citet{Cantiello2024} and assuming an intrinsic scatter of 0.19 mag, we obtain $D_{\rm SBF}=2.26 \pm 0.24$~Mpc. These distances are fully consistent with the fiducial distance but with a slightly higher distance uncertainty of $\sim 10\%$.

\begin{figure*}
    \centering
    \begin{minipage}{0.51\linewidth}
        \centering
        \includegraphics[width=\linewidth]{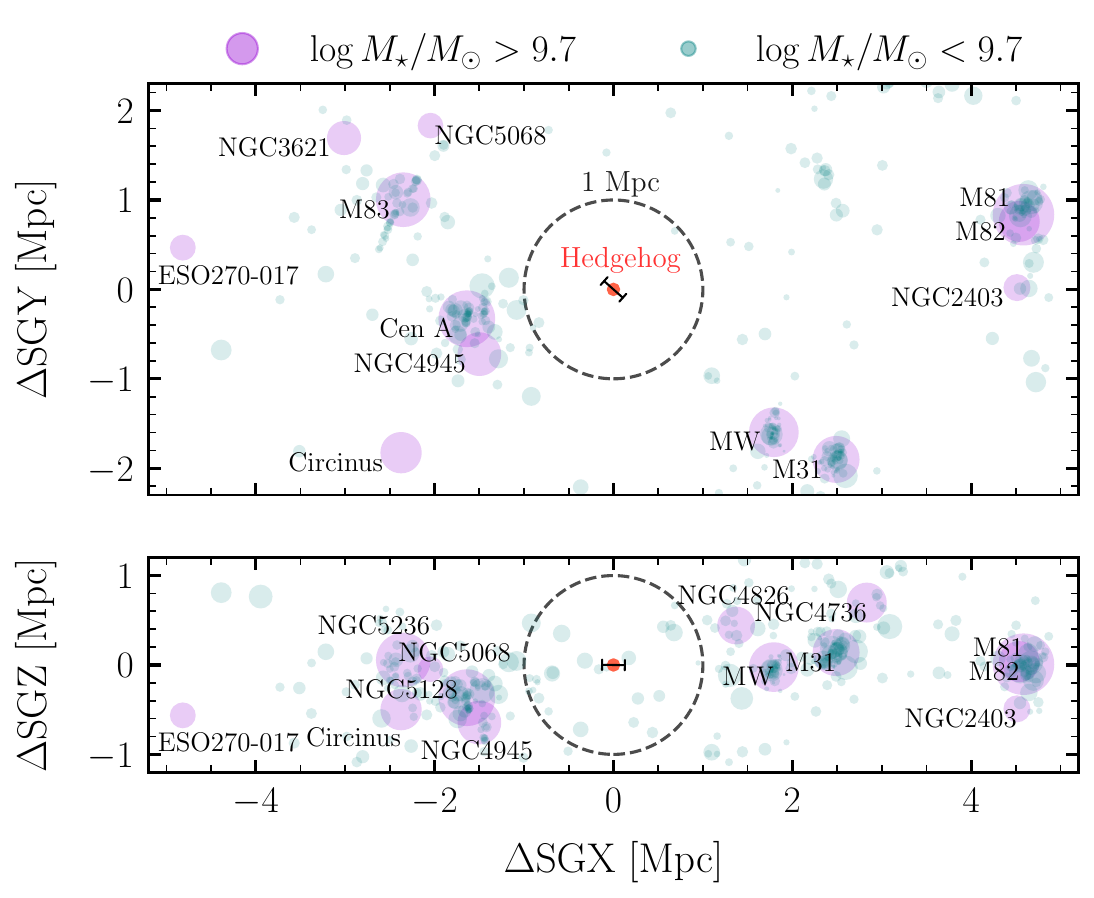}
    \end{minipage}
    \begin{minipage}{0.45\linewidth}
        \centering
            \includegraphics[width=\linewidth]{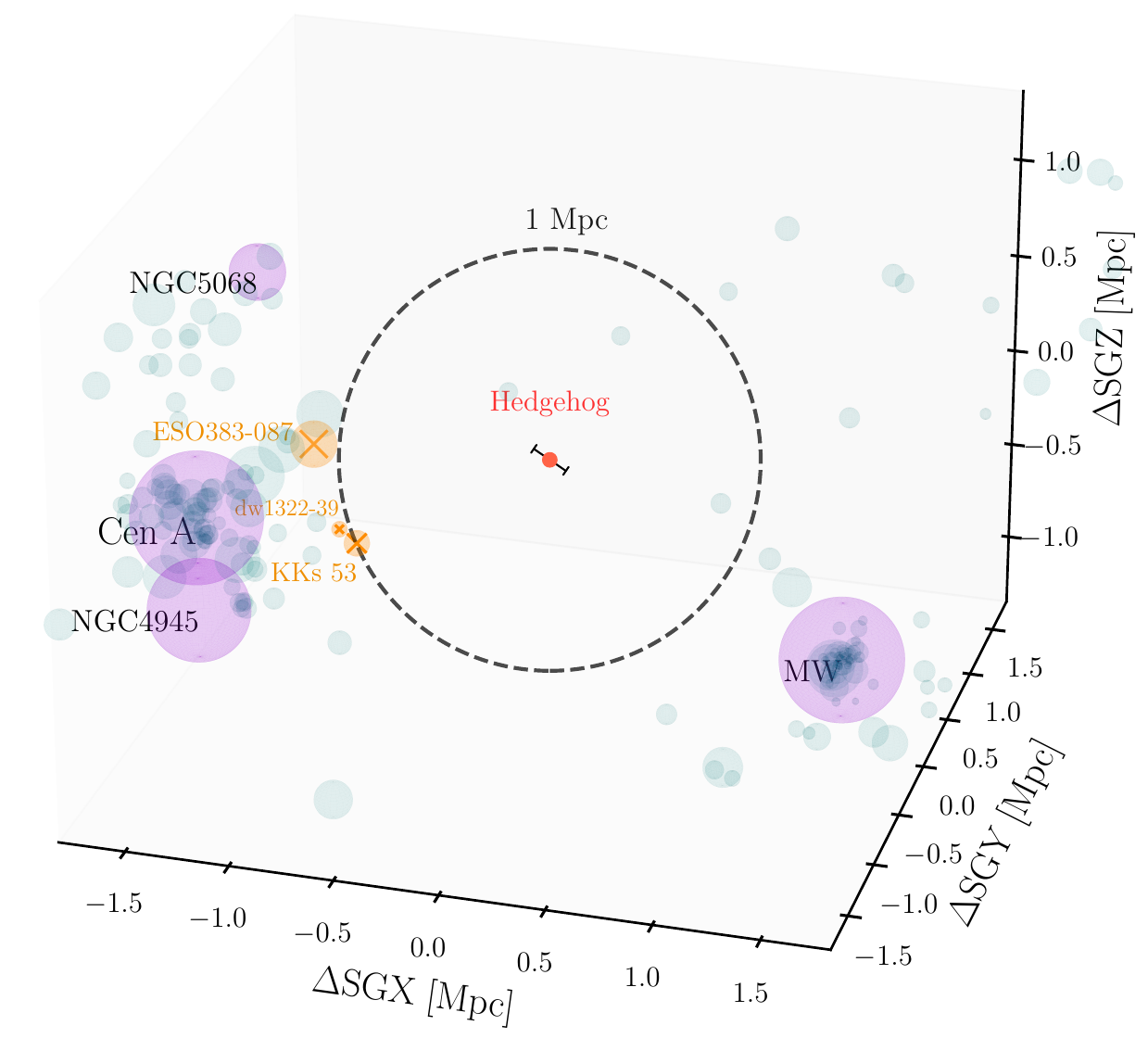}
    \end{minipage}
    \caption{Environment of Hedgehog. We show the distribution of neighboring galaxies from \citet{Karachentsev2013} in the supergalactic coordinate system centered at Hedgehog. The left panels show the 2D projected distributions and the right panel provides a close-up view of the environment in 3D. The size of the circle represents the virial radius of the host halo. Galaxies that are more (less) massive than $M_\star = 10^{9.7}\ M_\odot$ are shown in purple (blue). Hedgehog is 1.7~Mpc ($\sim 4-5\ R_{\rm vir}$) from its nearest galaxy group, Centaurus A. Hedgehog's closest neighbors, highlighted in orange, are at least 1~Mpc away. Hedgehog is thus one of the most isolated quiescent dwarf galaxies. An interactive version of this figure is available at the author's website, \url{https://astrojacobli.github.io/research/Hedgehog/}. An interactive version of this figure is also available in the \href{online article}{https://doi.org/10.3847/2041-8213/ad5b59}.}
    \label{fig:environment}
\end{figure*}

\section{Environment}\label{sec:environment}

We explore the environment of Hedgehog by showing its 3D location with respect to the known galaxies. We take the galaxies in the Updated Nearby Galaxy Catalog \citep[][]{Karachentsev2013} and calculate their coordinates in the supergalactic coordinate system. The stellar mass is estimated by taking the $K_s$-band luminosity in \citet{Karachentsev2013} and assuming a mass-to-light ratio of $M_\star / L_{\rm K_s} = 1$ \citep{Bell2003}\footnote{Other authors use a lower mass-to-light ratio than we use here \citep[e.g.,][]{CarlstenELVES2022}; thus, the stellar mass and virial radius derived in this work should be considered as upper limits.}. We also calculate the virial radius using \code{SatGen} \citep{Jiang2019} by assuming a stellar-to-halo mass relation from \citet{Rodriguez-Puebla2017}.

The left panels of Figure \ref{fig:environment} show the $XY$ and $XZ$ projections of the 3D distribution of nearby galaxies. We plot the locations of galaxies relative to Hedgehog for convenience. Hedgehog is shown as a red dot, with the error bar indicating the projected $1\sigma$ distance uncertainty. Galaxies that are more massive than $\log M_\star/M_\odot > 9.7$ ($\sim10\%$ of the stellar mass of the MW) are shown as purple circles, with the circle size corresponding to the projected virial radius. The fainter blue dots are galaxies with $\log M_\star/M_\odot < 9.7$. The dashed circle around Hedgehog indicates a sphere with a radius of 1 Mpc. Hedgehog lives in a low-density environment with no massive neighboring galaxies. The galaxy groups closest to Hedgehog are the Centaurus A (Cen~A) group, the Local Group, and the M83 group. Hedgehog is 1.7~Mpc from Cen~A, 2.4 Mpc from the MW, and 2.55~Mpc from M83 and thus is classified as a field dwarf by the criteria in \citet{Geha2012}. The Cen~A group has a virial mass ranging from $3.6\times 10^{12}\ M_\odot$ \citep{Dumont2024} and $5.3\times10^{12}\ M_\odot$ \citep{Muller2022} to $8\times 10^{12}\ M_\odot$ \citep{Karachentsev2007}, being about $4-8$ times more massive than the MW \citep{Patel2018}. Equivalently, Cen~A has a virial radius of $R_{\rm vir}\approx 320-410\ \mathrm{kpc}$. Hedgehog is $4-5\ R_{\rm vir}$ away from the Cen~A group, placing it in a very isolated environment.

We show a close-up view of the 3D environment in the right panel of Figure \ref{fig:environment}. Similar to the left panels, the sphere's size also represents the halo's virial radius. The Cen~A group is shown together with the MW group. We highlight the three closest neighbors of Hedgehog in orange. The closest neighbor of Hedgehog, KKs~53 (\citealt{Huchtmeier2001}; also known as [KK2000]~53 in \citealt{Karachentsev2013} and Cen~7 in \citealt{Muller2017}), is 1.00~Mpc away from Hedgehog with a distance of $D_{\rm TRGB}=2.93$~Mpc \citep{Tully2015} and a $K_s$-band luminosity of $\log L_{\rm K_s} = 7.46$ \citep{Karachentsev2013}. The second-closest neighbor is dw1322-39, a dwarf irregular at $D_{\rm TRGB} = 2.95$~Mpc \citep{Muller2019} and $\log L_{\rm K_s} = 6.12$ \citep{Karachentsev2013}. It is 1.03~Mpc away from Hedgehog and is considered a satellite of the Cen A group \citep{Muller2019}. The third-closest is ESO383--087, located 1.10~Mpc away from Hedgehog. It is a blue dwarf galaxy at $D_{\rm TRGB}=3.19$~Mpc with $\log L_{\rm K_s} = 9.05$ \citep{Karachentsev2013}. Hedgehog's neighbors are mostly low-mass dwarf galaxies distributed toward the Cen~A group. We do not find any neighboring galaxies within 1~Mpc from Hedgehog. This makes Hedgehog one of the most isolated quiescent dwarf galaxies in the Local Volume. As a comparison, Tucana B, whose closest neighbor (IC~5152) is 620 kpc away, is less isolated than Hedgehog. KKs~3 and KKR~25 have a similar degree of isolation as Hedgehog.

\section{Structure and Stellar Population}\label{sec:sp}

\subsection{Luminosity and Size of Hedgehog}
We calculate the physical properties of Hedgehog and list them in Table \ref{tab:property}. Hedgehog has an effective radius of $r_{\rm eff} = 176 \pm 14\ \rm{pc}$, and a $g$-band absolute magnitude of $M_g = -9.56 \pm 0.14$ mag. To compare with the literature, we convert our measured photometry to the $V$ band using the transformation $V = g - 0.5784 \cdot (g-r) - 0.0038$ \footnote{\url{https://www.sdss3.org/dr8/algorithms/sdssUBVRITransform.php\#Lupton2005}}. We also estimate the stellar mass of Hedgehog using the color--$M_\star/L$ relation from \citet{Into2013}, which uses a \citet{Kroupa2001} initial mass function. The derived stellar mass is $M_\star = 10^{5.8\pm0.1}\, M_\odot$ when using the $g-i$ color, and $M_\star = 10^{5.9\pm0.1}\, M_\odot$ when using the $g-r$ color. We therefore quote the stellar mass as $M_\star = 10^{5.8 \pm 0.2}\ M_\odot$, considering the uncertainties in photometry, color, and the color--$M_\star/L$ relation. 

\begin{figure}
    \centering 
    \includegraphics[width=1.\linewidth]{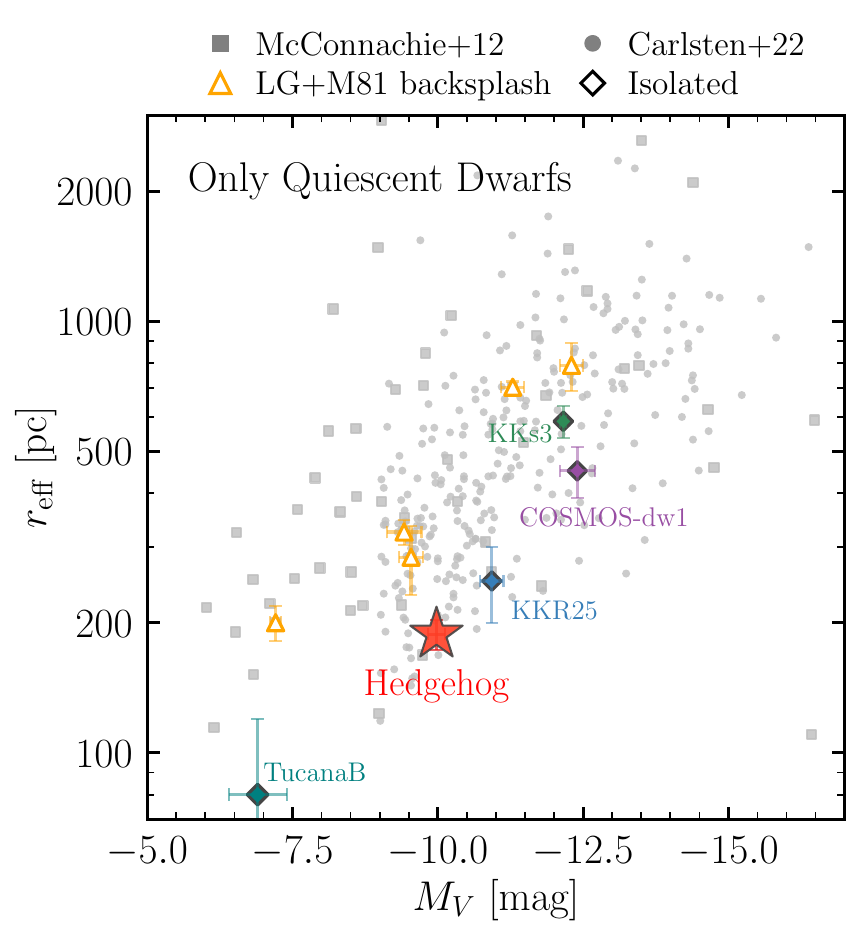}
    \caption{The $V$-band absolute magnitude and size of Hedgehog, together with other isolated quiescent dwarf galaxies (KKR 25, \citealt{Makarov2012}; KKs 3, \citealt{Karachentsev2015_KKs3}; COSMOS-dw1, \citealt{Xi2018,Polzin2021}; Tucana B, \citealt{Sand2022}) and potential backsplash galaxies in the Local Group and M81 (Cetus, Tucana, And XVIII, Eri II, and Blobby; shown in yellow triangle). The quiescent dwarfs at $D<2$~Mpc \citep{McConnachie2012} and early-type satellites of MW analogs \citep{CarlstenELVES2022} are shown in gray for reference. Hedgehog agrees with the size--luminosity relation but is on the smaller side. }
    \label{fig:mass-size}
\end{figure}

To place Hedgehog in a broader context, we plot its $V$-band absolute magnitude and effective radius in Figure \ref{fig:mass-size}, together with a few other isolated quiescent dwarf galaxies from the literature. We also show the dwarfs that are considered backsplash satellites of MW and M31 as yellow triangles (including Cetus, Tucana, And~XVIII, Eri~II; \citealt{Buck2019,Blana2020,SantosSantos2023}). For reference, we also show the quiescent dwarf galaxies within $D<2$~Mpc from \citet{McConnachie2012}\footnote{\url{https://www.cadc-ccda.hia-iha.nrc-cnrc.gc.ca/en/community/nearby/}} and \citet{Putman2021}, as well as the early-type satellites of MW analogs at $D<12$~Mpc from \citet{CarlstenELVES2022}. Compared with LG and LV quenched dwarf galaxies, Hedgehog seems to have a smaller size for its luminosity but is within the scatter of the luminosity--size relation. We also compare the size of Hedgehog with the mass--size relations from \citet[][based on LG dwarfs]{Danieli2018} and \citet[][based on LV satellites]{ELVES-I}. Hedgehog sits below both of the average mass--size relations but within the 1.5$\sigma$ scatter. Therefore, we conclude that Hedgehog is consistent with the mass--size relation of LG and LV dwarfs. 

Intriguingly, most isolated quiescent dwarfs (highlighted as diamonds in Figure \ref{fig:mass-size}) exhibit smaller sizes compared to the LG and LV quiescent dwarfs. Given that all of these galaxies were discovered serendipitously, their smaller sizes may result from observational bias: for a given luminosity, smaller galaxies have higher surface brightnesses and are therefore easier to identify. We discuss other physical interpretations in \S\ref{sec:discuss}. A homogeneous search for field dwarf galaxies in the Local Volume would provide a more objective assessment of this issue. 

\subsection{Stellar Population}
As described in \S\ref{sec:galex}, we do not detect significant UV emission from Hedgehog, implying that Hedgehog is a quiescent galaxy. The lack of UV emission indicates the absence of recent star formation on a timescale of $<100$~Myr \citep{Lee2011}. To be more precise, we calculate the upper limits of UV-based star formation rates (SFRs) following Equation 3 in \citet{Iglesias2006}. Similar to \citet{Karunakaran2021}, we do not attempt to correct the internal dust extinction because that would require infrared data and also because the dust in dwarf galaxies is likely negligible. The SFR of Hedgehog has a $2\sigma$ upper limit of $\rm SFR_{\rm UV} \lesssim 10^{-4}\ M_\odot\, \rm{yr}^{-1}$. This limit is similar to or lower than the UV-detected satellites of MW analogs \citep{Karunakaran2021,Karunakaran2022}. Because the \textit{GALEX} data used here are relatively shallower than the data used in \citet{Karunakaran2022}, it is possible that Hedgehog has a much lower SFR than $10^{-4}\ M_\odot\, \rm{yr}^{-1}$.

The integrated color of a dwarf galaxy is often used as a proxy for its stellar population, and the morphology can also roughly represent the stellar population, as early-type dwarfs are typically red and quenched \citep{ELVES-I,Greene2023}. Hedgehog has a red color ($g-r=0.49$, $g-i=0.62$) and a symmetric smooth morphology without star-forming regions or dust lanes, being consistent with a typical dwarf spheroidal galaxy.

To better understand the stellar population, we generate simple stellar populations (SSPs) with various ages and metallicities using the MIST isochrones \citep{Dotter2016,Choi2016} and \code{ArtPop} \citep{artpop} and compare the corresponding colors with the measured colors \citep{Casey2023}. Because of the degeneracy between age and metallicity, we anchor the metallicity of Hedgehog to the metallicity predicted from the mass--metallicity relation from \citet{Kirby2013}: $\feh = -1.75 \pm 0.17$. An SSP at this metallicity and with age $t_{\rm age} = 5-7\ \rm{Gyr}$ is consistent with the colors of Hedgehog within the measurement uncertainty. We further validate the inferred stellar population and distance in Appendix \ref{ap:artpop} by generating mock galaxy images using \code{ArtPop}. However, we cannot rule out the presence of an intermediate-age stellar population ($\sim ~0.2-1$~Gyr) with the optical and UV data alone. Deep space-based observation is needed to measure the CMD and derive the star formation history with better temporal resolution \citep[e.g.,][]{Makarov2012,Weisz2011,Weisz2014}. 

We also search for neutral gas at the location of Hedgehog and do not find a coincident \ion{H}{1} detection in the \ion{H}{1} Parkes All-Sky Survey \citep[HIPASS;][]{Barnes2001,Kalberla2015} and Galactic All-Sky Survey \citep[GASS;][]{McClure-Griffiths2009} data. GASS covers a radial velocity range of $-400\ \mathrm{km\ s}^{-1}$ to $+500\ \mathrm{km\ s}^{-1}$ with a higher resolution in velocity, whereas HIPASS covers range of $-1280\ \mathrm{km\ s}^{-1}$ to $+12,700\ \mathrm{km\ s}^{-1}$ and is more sensitive than GASS by a factor of 4. However, the radial velocity of Hedgehog is currently unknown. If it is a backsplash galaxy from the relatively nearby Cen~A group, it might have a similar velocity to Cen~A ($\sim 560\ \mathrm{km\ s}^{-1}$, \citealt{Koss2022}). Therefore, the non-detection in GASS might be due to its inadequate velocity range. Using HIPASS, we derive the $3\sigma$ upper limit of \ion{H}{1} mass to be $M_{\rm HI} < 10^{6.0}\ M_\odot$, assuming an RMS noise level of 13 mJy and a velocity width of $20\ \mathrm{km\ s^{-1}}$. The upcoming WALLABY survey \citep{WALLABY2020}, which has a detection limit of $M_{\rm HI}\sim 10^{5}\ M_\odot$ at $D=2.4$~Mpc, will help us determine the neutral gas content of Hedgehog.

\section{Discussion}\label{sec:discuss}
Hedgehog is an isolated dwarf galaxy at a distance of $2.40\pm 0.15$~Mpc with a stellar mass $M_\star \approx 10^{5.8\pm 0.2}\ M_\odot$. It has no neighboring galaxies within 1~Mpc, and is 1.7~Mpc ($\sim 4-5\ R_{\rm vir}$) from Cen~A, its nearest galaxy group. Hedgehog is thus one of the most isolated quiescent dwarfs found to date and is less massive than the other extremely isolated dwarfs KKs~3 and KKR~25. It shows no sign of recent star formation and has an old stellar population. The challenge is to understand how its star formation is quenched given its isolation.

The most straightforward explanation is that Hedgehog is a backsplash galaxy from the nearest galaxy group. Simulations of satellites around MW-like groups have shown that backsplash galaxies contribute 13\% of the dwarf galaxies found between $R_{\rm vir}$ and 1.5 Mpc \citep{Teyssier2012}, and the fraction could be as high as $\sim 50$\% between $R_{\rm vir}$ and $2.5 R_{\rm vir}$ \citep{Buck2019}. \citet{Benavides2021} even identified a handful of backsplash ultra-diffuse galaxies out to $3-4\ R_{\rm vir}$. Similar to satellite galaxies, backsplash galaxies undergo quenching through ram pressure stripping and tidal interactions, resulting in a quenched fraction comparable to that of the satellites \citep{Simpson2018}.

In the backsplash scenario, Hedgehog is most likely to originate from the Cen~A group given their proximity. However, its distance to Cen~A might be too far ($\sim 4-5\ R_{\rm vir}$) compared with simulations. Given its current distance of 1.7~Mpc from Cen A, Hedgehog's travel time, assuming an ejection velocity of $v_{\rm esp} \approx \sqrt{2} \ v_{\rm circ}$ with $v_{\rm circ} \approx 260\pm 60\ \rm{km\ s^{-1}}$ \citep{Muller2022}, would be 4--6 Gyr. If Hedgehog was quenched within 1--2~Gyr after it had the pericenter passage in the Cen~A group \citep{Slater2014,Wetzel2015,Simpson2018,Greene2023}, the estimated travel time agrees quite well with the inferred age of Hedgehog's stellar population ($t_{\rm age}\approx 5-7$~Gyr). However, it is important to note that the inferred age of Hedgehog is still quite uncertain due to the age--metallicity degeneracy. Additionally, if a major merger of Cen A occurred around 2 Gyr ago \citep{Wang2020}, it would complicate this timing argument.


On the other hand, it might not be surprising to find quiescent field dwarfs with $M_\star \approx 10^{6}\ M_\odot$. Despite that most isolated dwarfs with $M_\star = 10^{7-9}\ M_\odot$ are star-forming \citep{Geha2012}, field dwarf galaxies have a quenched fraction of $\sim 20\%$ at $M_\star \approx 10^{6-7}\ M_\odot$ \citep{Slater2014}. Using the empirical galaxy-halo model \textsc{UniverseMachine} \citep{Behroozi2019}, \citet{Wang2024} predicted a quenched fraction of $\sim 30\%$ for field dwarfs with $M_\star \approx 10^{6.5}\ M_\odot$. In hydrodynamical zoom-in simulations, the quenched fraction ranges from $\sim 25\%$ \citep{Christensen2024} to $\sim 70\%$ \citep{Simpson2018} in this mass range.

Several physical mechanisms are invoked to explain quenching in the field. Field dwarfs can have their gas stripped by the ram pressure from filament gas as they move through the cosmic web \citep{Benitez-Llambay2013, Simpson2018}. Stellar feedback can contribute to quenching \citep{Samuel2022}, although it typically results in only temporary quiescence and is unlikely to be the dominant mechanism for field dwarfs. Reionization, on the other hand, is the most likely cause of quenching for dwarfs with $M_\star \lesssim 10^{6}\ M_\odot$ \citep[e.g.,][]{Bullock2001,Ricotti2005}, and isolated quenched dwarfs are the best candidate reionization fossils \citep{Weisz2014}. The UV background from reionization could blow out the gas via photoevaporation \citep{Rees1986,Shapiro2004}, prolong gas cooling time by photoionization heating \citep{Efstathiou1992}, and prevent gas inflow, leading to quenching via starvation \citep{Okamoto2008,Katz2020}. Hedgehog's mass is close to the upper limit where reionization can effectively quench star formation. Recent simulations demonstrate that field dwarfs with $M_\star \approx 10^{5-6}\ M_\odot$ can be quenched by reionization, but can also be rejuvenated later \citep{Rey2020}. To remain quiescent, it is possible that Hedgehog has a slightly lower halo mass as indicated by its small half-light radius \citep{Kravtsov2013}, or experienced a stronger UV background due to the inhomogeneous nature of reionization. The puzzle of Hedgehog's formation and quenching highlights the complexity of galaxy evolution in the low-mass regime.

The size of isolated quiescent dwarf galaxies may provide an additional clue about their formation and quenching mechanisms. We notice that many isolated quiescent dwarfs have smaller sizes than the average luminosity--size relation in Figure \ref{fig:mass-size}. For backsplash galaxies, they would have very eccentric orbits and thus could be tidally heated during the pericenter passage \citep{Carleton2019} and puffed up in size \citep{Jiang2019}. If most isolated quiescent dwarfs are backsplash galaxies, they will exhibit larger sizes, which contradicts Figure \ref{fig:mass-size}. However, if they are reionization fossils, they will stop growing after being quenched at high redshift and thus have smaller sizes because the dark matter halos at high redshifts are smaller and more concentrated \citep{Manwadkar2022}. 


For Hedgehog, while the integrated colors alone do not rule out the presence of an intermediate-age stellar population, deep CMDs will enable us to uncover its detailed star formation history \citep[e.g.,][]{McQuinn2024} and test the reionization quenching scenario. Additionally, \citet{Buck2019} show that backsplash galaxies exhibit lower velocity dispersion compared to field dwarfs. Therefore, radial velocity and velocity dispersion measurements will be crucial in confirming or ruling out the backsplash scenario. With the upcoming Vera C. Rubin Observatory’s Legacy Survey of Space and Time \citep{LSST2019}, it will be possible to conduct a blind search for field dwarf galaxies out to $\sim 20$~Mpc with SBF in a homogeneous way and better characterize quenching as a function of environment and mass.


\section*{Acknowledgment}
We thank Ava Polzin for curating a list of isolated quiescent dwarf galaxies in the literature\footnote{\url{https://avapolzin.github.io/projects/quench_list/}}. We thank the anonymous reviewer for the constructive suggestions, which made this manuscript clearer. J.L. is grateful for the discussions with Yifei Luo, Shun Wang, Burçin Mutlu-Pakdil, Michele Cantiello, Oliver M\"uller, Yunchong Wang, Yilun Ma, and Sihao Cheng. J.L. thanks Khalil Fong for his music. J.E.G. gratefully acknowledges support from NSF grant NSF AAG/AWD1007052.

This paper includes data gathered with the 6.5 m Magellan Telescopes located at Las Campanas Observatory, Chile. 

The Legacy Surveys consist of three individual and complementary projects: the Dark Energy Camera Legacy Survey (DECaLS; Proposal ID \#2014B-0404; PIs: David Schlegel and Arjun Dey), the Beijing-Arizona Sky Survey (BASS; NOAO Prop. ID \#2015A-0801; PIs: Zhou Xu and Xiaohui Fan), and the Mayall z-band Legacy Survey (MzLS; Prop. ID \#2016A-0453; PI: Arjun Dey). DECaLS, BASS and MzLS together include data obtained, respectively, at the Blanco telescope, Cerro Tololo Inter-American Observatory, NSF’s NOIRLab; the Bok telescope, Steward Observatory, University of Arizona; and the Mayall telescope, Kitt Peak National Observatory, NOIRLab. Pipeline processing and analyses of the data were supported by NOIRLab and the Lawrence Berkeley National Laboratory (LBNL). The Legacy Surveys project is honored to be permitted to conduct astronomical research on Iolkam Du’ag (Kitt Peak), a mountain with particular significance to the Tohono O’odham Nation.

NOIRLab is operated by the Association of Universities for Research in Astronomy (AURA) under a cooperative agreement with the National Science Foundation. LBNL is managed by the Regents of the University of California under contract to the U.S. Department of Energy.

This project used data obtained with the Dark Energy Camera (DECam), which was constructed by the Dark Energy Survey (DES) collaboration. Funding for the DES Projects has been provided by the U.S. Department of Energy, the U.S. National Science Foundation, the Ministry of Science and Education of Spain, the Science and Technology Facilities Council of the United Kingdom, the Higher Education Funding Council for England, the National Center for Supercomputing Applications at the University of Illinois at Urbana-Champaign, the Kavli Institute of Cosmological Physics at the University of Chicago, Center for Cosmology and Astro-Particle Physics at the Ohio State University, the Mitchell Institute for Fundamental Physics and Astronomy at Texas A\&M University, Financiadora de Estudos e Projetos, Fundacao Carlos Chagas Filho de Amparo, Financiadora de Estudos e Projetos, Fundacao Carlos Chagas Filho de Amparo a Pesquisa do Estado do Rio de Janeiro, Conselho Nacional de Desenvolvimento Cientifico e Tecnologico and the Ministerio da Ciencia, Tecnologia e Inovacao, the Deutsche Forschungsgemeinschaft and the Collaborating Institutions in the Dark Energy Survey. The Collaborating Institutions are Argonne National Laboratory, the University of California at Santa Cruz, the University of Cambridge, Centro de Investigaciones Energeticas, Medioambientales y Tecnologicas-Madrid, the University of Chicago, University College London, the DES-Brazil Consortium, the University of Edinburgh, the Eidgenossische Technische Hochschule (ETH) Zurich, Fermi National Accelerator Laboratory, the University of Illinois at Urbana-Champaign, the Institut de Ciencies de l’Espai (IEEC/CSIC), the Institut de Fisica d’Altes Energies, Lawrence Berkeley National Laboratory, the Ludwig Maximilians Universitat Munchen and the associated Excellence Cluster Universe, the University of Michigan, NSF’s NOIRLab, the University of Nottingham, the Ohio State University, the University of Pennsylvania, the University of Portsmouth, SLAC National Accelerator Laboratory, Stanford University, the University of Sussex, and Texas A\&M University.

BASS is a key project of the Telescope Access Program (TAP), which has been funded by the National Astronomical Observatories of China, the Chinese Academy of Sciences (the Strategic Priority Research Program “The Emergence of Cosmological Structures” Grant \# XDB09000000), and the Special Fund for Astronomy from the Ministry of Finance. The BASS is also supported by the External Cooperation Program of the Chinese Academy of Sciences (Grant \# 114A11KYSB20160057), and the Chinese National Natural Science Foundation (Grant \# 12120101003, \# 11433005).

The Legacy Survey team makes use of data products from the Near-Earth Object Wide-field Infrared Survey Explorer (NEOWISE), which is a project of the Jet Propulsion Laboratory/California Institute of Technology. NEOWISE is funded by the National Aeronautics and Space Administration.

The Legacy Surveys imaging of the DESI footprint is supported by the Director, Office of Science, Office of High Energy Physics of the U.S. Department of Energy under Contract No. DE-AC02-05CH1123, by the National Energy Research Scientific Computing Center, a DOE Office of Science User Facility under the same contract; and by the U.S. National Science Foundation, Division of Astronomical Sciences under Contract No. AST-0950945 to NOAO.

This research used observations made with the NASA Galaxy Evolution Explorer. GALEX is operated for NASA by the California Institute of Technology under NASA contract NAS5-98034.

This research has used the NASA/IPAC Extragalactic Database (NED), which is funded by the National Aeronautics and Space Administration and operated by the California Institute of Technology. This research has also used the NASA/IPAC Infrared Science Archive, which is funded by the National Aeronautics and Space Administration and operated by the California Institute of Technology.

The authors are pleased to acknowledge that the work reported in this paper was substantially performed using the Princeton Research Computing resources at Princeton University, a consortium of groups led by the Princeton Institute for Computational Science and Engineering (PICSciE) and the Office of Information Technology's Research Computing.

\vspace{1em}
\facilities{Magellan:Baade (IMACS), Blanco (DECam), \textit{GALEX}.}

\software{\href{http://www.numpy.org}{\code{NumPy}} \citep{Numpy},
          \href{https://www.astropy.org/}{\code{Astropy}} 
          \citep{astropy:2013, astropy:2018, astropy:2022},
          \href{https://www.scipy.org}{\code{SciPy}} \citep{scipy}, \href{https://matplotlib.org}{\code{Matplotlib}} \citep{matplotlib},
          \href{https://artpop.readthedocs.io/en/latest/index.html}{\code{ArtPop}} \citep{Greco2021}, \href{https://photutils.readthedocs.io/en/stable/}{\code{photutils}} \citep{photutils}, \href{https://www.astromatic.net/software/sextractor/}{\code{SExtractor}} \citep{Bertin1996},
          \href{https://www.astromatic.net/software/swarp/}{\code{SWarp}} \citep{swarp},
          \href{https://www.astromatic.net/software/scamp/}{\code{SCAMP}} \citep{scamp},
          \href{https://astrometry.net/}{\code{astrometry.net}} \citep{astrometry_net},
          \href{https://www.astromatic.net/software/psfex/}{\code{PSFEx}} \citep{psfex},
          \href{https://sep.readthedocs.io/en/v1.1.x/}{\code{sep}} \citep{Barbary2016},
          \href{https://www.mpe.mpg.de/~erwin/code/imfit/}{\code{imfit}} \citep{imfit},
          \href{https://github.com/kbarbary/sfdmap}{\code{sfdmap}}.
          }

\bibliography{cleaned_citation,software}

\begin{thebibliography}{}
\expandafter\ifx\csname natexlab\endcsname\relax\def\natexlab#1{#1}\fi
\providecommand{\url}[1]{\href{#1}{#1}}
\providecommand{\dodoi}[1]{doi:~\href{http://doi.org/#1}{\nolinkurl{#1}}}
\providecommand{\doeprint}[1]{\href{http://ascl.net/#1}{\nolinkurl{http://ascl.net/#1}}}
\providecommand{\doarXiv}[1]{\href{https://arxiv.org/abs/#1}{\nolinkurl{https://arxiv.org/abs/#1}}}

\bibitem[{{Applebaum} {et~al.}(2021){Applebaum}, {Brooks}, {Christensen}, {Munshi}, {Quinn}, {Shen}, \& {Tremmel}}]{Applebaum2021}
{Applebaum}, E., {Brooks}, A.~M., {Christensen}, C.~R., {et~al.} 2021, \apj, 906, 96, \dodoi{10.3847/1538-4357/abcafa}

\bibitem[{{Astropy Collaboration} {et~al.}(2013){Astropy Collaboration}, {Robitaille}, {Tollerud}, {Greenfield}, {Droettboom}, {Bray}, {Aldcroft}, {Davis}, {Ginsburg}, {Price-Whelan}, {Kerzendorf}, {Conley}, {Crighton}, {Barbary}, {Muna}, {Ferguson}, {Grollier}, {Parikh}, {Nair}, {Unther}, {Deil}, {Woillez}, {Conseil}, {Kramer}, {Turner}, {Singer}, {Fox}, {Weaver}, {Zabalza}, {Edwards}, {Azalee Bostroem}, {Burke}, {Casey}, {Crawford}, {Dencheva}, {Ely}, {Jenness}, {Labrie}, {Lim}, {Pierfederici}, {Pontzen}, {Ptak}, {Refsdal}, {Servillat}, \& {Streicher}}]{astropy:2013}
{Astropy Collaboration}, {Robitaille}, T.~P., {Tollerud}, E.~J., {et~al.} 2013, \aap, 558, A33, \dodoi{10.1051/0004-6361/201322068}

\bibitem[{{Astropy Collaboration} {et~al.}(2018){Astropy Collaboration}, {Price-Whelan}, {Sip{\H{o}}cz}, {G{\"u}nther}, {Lim}, {Crawford}, {Conseil}, {Shupe}, {Craig}, {Dencheva}, {Ginsburg}, {Vand erPlas}, {Bradley}, {P{\'e}rez-Su{\'a}rez}, {de Val-Borro}, {Aldcroft}, {Cruz}, {Robitaille}, {Tollerud}, {Ardelean}, {Babej}, {Bach}, {Bachetti}, {Bakanov}, {Bamford}, {Barentsen}, {Barmby}, {Baumbach}, {Berry}, {Biscani}, {Boquien}, {Bostroem}, {Bouma}, {Brammer}, {Bray}, {Breytenbach}, {Buddelmeijer}, {Burke}, {Calderone}, {Cano Rodr{\'\i}guez}, {Cara}, {Cardoso}, {Cheedella}, {Copin}, {Corrales}, {Crichton}, {D'Avella}, {Deil}, {Depagne}, {Dietrich}, {Donath}, {Droettboom}, {Earl}, {Erben}, {Fabbro}, {Ferreira}, {Finethy}, {Fox}, {Garrison}, {Gibbons}, {Goldstein}, {Gommers}, {Greco}, {Greenfield}, {Groener}, {Grollier}, {Hagen}, {Hirst}, {Homeier}, {Horton}, {Hosseinzadeh}, {Hu}, {Hunkeler}, {Ivezi{\'c}}, {Jain}, {Jenness}, {Kanarek}, {Kendrew}, {Kern}, {Kerzendorf}, {Khvalko}, {King}, {Kirkby}, {Kulkarni},
  {Kumar}, {Lee}, {Lenz}, {Littlefair}, {Ma}, {Macleod}, {Mastropietro}, {McCully}, {Montagnac}, {Morris}, {Mueller}, {Mumford}, {Muna}, {Murphy}, {Nelson}, {Nguyen}, {Ninan}, {N{\"o}the}, {Ogaz}, {Oh}, {Parejko}, {Parley}, {Pascual}, {Patil}, {Patil}, {Plunkett}, {Prochaska}, {Rastogi}, {Reddy Janga}, {Sabater}, {Sakurikar}, {Seifert}, {Sherbert}, {Sherwood-Taylor}, {Shih}, {Sick}, {Silbiger}, {Singanamalla}, {Singer}, {Sladen}, {Sooley}, {Sornarajah}, {Streicher}, {Teuben}, {Thomas}, {Tremblay}, {Turner}, {Terr{\'o}n}, {van Kerkwijk}, {de la Vega}, {Watkins}, {Weaver}, {Whitmore}, {Woillez}, {Zabalza}, \& {Astropy Contributors}}]{astropy:2018}
{Astropy Collaboration}, {Price-Whelan}, A.~M., {Sip{\H{o}}cz}, B.~M., {et~al.} 2018, \aj, 156, 123, \dodoi{10.3847/1538-3881/aabc4f}

\bibitem[{{Astropy Collaboration} {et~al.}(2022){Astropy Collaboration}, {Price-Whelan}, {Lim}, {Earl}, {Starkman}, {Bradley}, {Shupe}, {Patil}, {Corrales}, {Brasseur}, {N{"o}the}, {Donath}, {Tollerud}, {Morris}, {Ginsburg}, {Vaher}, {Weaver}, {Tocknell}, {Jamieson}, {van Kerkwijk}, {Robitaille}, {Merry}, {Bachetti}, {G{"u}nther}, {Aldcroft}, {Alvarado-Montes}, {Archibald}, {B{'o}di}, {Bapat}, {Barentsen}, {Baz{'a}n}, {Biswas}, {Boquien}, {Burke}, {Cara}, {Cara}, {Conroy}, {Conseil}, {Craig}, {Cross}, {Cruz}, {D'Eugenio}, {Dencheva}, {Devillepoix}, {Dietrich}, {Eigenbrot}, {Erben}, {Ferreira}, {Foreman-Mackey}, {Fox}, {Freij}, {Garg}, {Geda}, {Glattly}, {Gondhalekar}, {Gordon}, {Grant}, {Greenfield}, {Groener}, {Guest}, {Gurovich}, {Handberg}, {Hart}, {Hatfield-Dodds}, {Homeier}, {Hosseinzadeh}, {Jenness}, {Jones}, {Joseph}, {Kalmbach}, {Karamehmetoglu}, {Ka{l}uszy{'n}ski}, {Kelley}, {Kern}, {Kerzendorf}, {Koch}, {Kulumani}, {Lee}, {Ly}, {Ma}, {MacBride}, {Maljaars}, {Muna}, {Murphy}, {Norman}, {O'Steen},
  {Oman}, {Pacifici}, {Pascual}, {Pascual-Granado}, {Patil}, {Perren}, {Pickering}, {Rastogi}, {Roulston}, {Ryan}, {Rykoff}, {Sabater}, {Sakurikar}, {Salgado}, {Sanghi}, {Saunders}, {Savchenko}, {Schwardt}, {Seifert-Eckert}, {Shih}, {Jain}, {Shukla}, {Sick}, {Simpson}, {Singanamalla}, {Singer}, {Singhal}, {Sinha}, {Sip{H{o}}cz}, {Spitler}, {Stansby}, {Streicher}, {{{S}}umak}, {Swinbank}, {Taranu}, {Tewary}, {Tremblay}, {Val-Borro}, {Van Kooten}, {Vasovi{'c}}, {Verma}, {de Miranda Cardoso}, {Williams}, {Wilson}, {Winkel}, {Wood-Vasey}, {Xue}, {Yoachim}, {Zhang}, {Zonca}, \& {Astropy Project Contributors}}]{astropy:2022}
{Astropy Collaboration}, {Price-Whelan}, A.~M., {Lim}, P.~L., {et~al.} 2022, \apj, 935, 167, \dodoi{10.3847/1538-4357/ac7c74}

\bibitem[{Barbary(2016)}]{Barbary2016}
Barbary, K. 2016, Journal of Open Source Software, 1(6), 58, \dodoi{10.21105/joss.0005}

\bibitem[{{Barnes} {et~al.}(2001){Barnes}, {Staveley-Smith}, {de Blok}, {Oosterloo}, {Stewart}, {Wright}, {Banks}, {Bhathal}, {Boyce}, {Calabretta}, {Disney}, {Drinkwater}, {Ekers}, {Freeman}, {Gibson}, {Green}, {Haynes}, {te Lintel Hekkert}, {Henning}, {Jerjen}, {Juraszek}, {Kesteven}, {Kilborn}, {Knezek}, {Koribalski}, {Kraan-Korteweg}, {Malin}, {Marquarding}, {Minchin}, {Mould}, {Price}, {Putman}, {Ryder}, {Sadler}, {Schr{\"o}der}, {Stootman}, {Webster}, {Wilson}, \& {Ye}}]{Barnes2001}
{Barnes}, D.~G., {Staveley-Smith}, L., {de Blok}, W.~J.~G., {et~al.} 2001, \mnras, 322, 486, \dodoi{10.1046/j.1365-8711.2001.04102.x}

\bibitem[{{Behroozi} {et~al.}(2019){Behroozi}, {Wechsler}, {Hearin}, \& {Conroy}}]{Behroozi2019}
{Behroozi}, P., {Wechsler}, R.~H., {Hearin}, A.~P., \& {Conroy}, C. 2019, \mnras, 488, 3143, \dodoi{10.1093/mnras/stz1182}

\bibitem[{{Bell} {et~al.}(2003){Bell}, {McIntosh}, {Katz}, \& {Weinberg}}]{Bell2003}
{Bell}, E.~F., {McIntosh}, D.~H., {Katz}, N., \& {Weinberg}, M.~D. 2003, \apjs, 149, 289, \dodoi{10.1086/378847}

\bibitem[{{Benavides} {et~al.}(2021){Benavides}, {Sales}, {Abadi}, {Pillepich}, {Nelson}, {Marinacci}, {Cooper}, {Pakmor}, {Torrey}, {Vogelsberger}, \& {Hernquist}}]{Benavides2021}
{Benavides}, J.~A., {Sales}, L.~V., {Abadi}, M.~G., {et~al.} 2021, Nature Astronomy, 5, 1255, \dodoi{10.1038/s41550-021-01458-1}

\bibitem[{{Ben{\'\i}tez-Llambay} {et~al.}(2013){Ben{\'\i}tez-Llambay}, {Navarro}, {Abadi}, {Gottl{\"o}ber}, {Yepes}, {Hoffman}, \& {Steinmetz}}]{Benitez-Llambay2013}
{Ben{\'\i}tez-Llambay}, A., {Navarro}, J.~F., {Abadi}, M.~G., {et~al.} 2013, \apjl, 763, L41, \dodoi{10.1088/2041-8205/763/2/L41}

\bibitem[{{Bertin}(2006)}]{scamp}
{Bertin}, E. 2006, in Astronomical Society of the Pacific Conference Series, Vol. 351, Astronomical Data Analysis Software and Systems XV, ed. C.~{Gabriel}, C.~{Arviset}, D.~{Ponz}, \& S.~{Enrique}, 112

\bibitem[{{Bertin}(2011)}]{psfex}
{Bertin}, E. 2011, in Astronomical Society of the Pacific Conference Series, Vol. 442, Astronomical Data Analysis Software and Systems XX, ed. I.~N. {Evans}, A.~{Accomazzi}, D.~J. {Mink}, \& A.~H. {Rots}, 435

\bibitem[{{Bertin} \& {Arnouts}(1996)}]{Bertin1996}
{Bertin}, E., \& {Arnouts}, S. 1996, \aaps, 117, 393, \dodoi{10.1051/aas:1996164}

\bibitem[{{Bertin} {et~al.}(2002){Bertin}, {Mellier}, {Radovich}, {Missonnier}, {Didelon}, \& {Morin}}]{swarp}
{Bertin}, E., {Mellier}, Y., {Radovich}, M., {et~al.} 2002, in Astronomical Society of the Pacific Conference Series, Vol. 281, Astronomical Data Analysis Software and Systems XI, ed. D.~A. {Bohlender}, D.~{Durand}, \& T.~H. {Handley}, 228

\bibitem[{{Bla{\~n}a} {et~al.}(2020){Bla{\~n}a}, {Burkert}, {Fellhauer}, {Schartmann}, \& {Alig}}]{Blana2020}
{Bla{\~n}a}, M., {Burkert}, A., {Fellhauer}, M., {Schartmann}, M., \& {Alig}, C. 2020, \mnras, 497, 3601, \dodoi{10.1093/mnras/staa2153}

\bibitem[{{Bradley} {et~al.}(2016){Bradley}, {Sipocz}, {Robitaille}, {Tollerud}, {Deil}, {Vin{\'\i}cius}, {Barbary}, {G{\"u}nther}, {Bostroem}, {Droettboom}, {Bray}, {Bratholm}, {Pickering}, {Craig}, {Pascual}, {Greco}, {Donath}, {Kerzendorf}, {Littlefair}, {Barentsen}, {D'Eugenio}, \& {Weaver}}]{photutils}
{Bradley}, L., {Sipocz}, B., {Robitaille}, T., {et~al.} 2016, {Photutils: Photometry tools}, Astrophysics Source Code Library, record ascl:1609.011

\bibitem[{{Buck} {et~al.}(2019){Buck}, {Macci{\`o}}, {Dutton}, {Obreja}, \& {Frings}}]{Buck2019}
{Buck}, T., {Macci{\`o}}, A.~V., {Dutton}, A.~A., {Obreja}, A., \& {Frings}, J. 2019, \mnras, 483, 1314, \dodoi{10.1093/mnras/sty2913}

\bibitem[{{Bullock} {et~al.}(2001){Bullock}, {Kolatt}, {Sigad}, {Somerville}, {Kravtsov}, {Klypin}, {Primack}, \& {Dekel}}]{Bullock2001}
{Bullock}, J.~S., {Kolatt}, T.~S., {Sigad}, Y., {et~al.} 2001, \mnras, 321, 559, \dodoi{10.1046/j.1365-8711.2001.04068.x}

\bibitem[{{Bullock} {et~al.}(2000){Bullock}, {Kravtsov}, \& {Weinberg}}]{Bullock2000}
{Bullock}, J.~S., {Kravtsov}, A.~V., \& {Weinberg}, D.~H. 2000, \apj, 539, 517, \dodoi{10.1086/309279}

\bibitem[{{Cantiello} {et~al.}(2005){Cantiello}, {Blakeslee}, {Raimondo}, {Mei}, {Brocato}, \& {Capaccioli}}]{Cantiello2005}
{Cantiello}, M., {Blakeslee}, J.~P., {Raimondo}, G., {et~al.} 2005, \apj, 634, 239, \dodoi{10.1086/491694}

\bibitem[{{Cantiello} {et~al.}(2024){Cantiello}, {Blakeslee}, {Ferrarese}, {C{\^o}t{\'e}}, {Raimondo}, {Cuillandre}, {Durrell}, {Gwyn}, {Hazra}, {Peng}, {Roediger}, {S{\'a}nchez-Janssen}, \& {Kurzner}}]{Cantiello2024}
{Cantiello}, M., {Blakeslee}, J.~P., {Ferrarese}, L., {et~al.} 2024, \apj, 966, 145, \dodoi{10.3847/1538-4357/ad3453}

\bibitem[{{Carleton} {et~al.}(2019){Carleton}, {Errani}, {Cooper}, {Kaplinghat}, {Pe{\~n}arrubia}, \& {Guo}}]{Carleton2019}
{Carleton}, T., {Errani}, R., {Cooper}, M., {et~al.} 2019, \mnras, 485, 382, \dodoi{10.1093/mnras/stz383}

\bibitem[{{Carleton} {et~al.}(2024{\natexlab{a}}){Carleton}, {Ellsworth-Bowers}, {Windhorst}, {Cohen}, {Conselice}, {Diego}, {Zitrin}, {Archer}, {McIntyre}, {Kamieneski}, {Jansen}, {Summers}, {D'Silva}, {Koekemoer}, {Coe}, {Driver}, {Frye}, {Grogin}, {Marshall}, {Nonino}, {Pirzkal}, {Robotham}, {Ryan}, {Ortiz}, {Tompkins}, {Willmer}, {Yan}, \& {Holwerda}}]{Carleton2024}
{Carleton}, T., {Ellsworth-Bowers}, T., {Windhorst}, R.~A., {et~al.} 2024{\natexlab{a}}, \apjl, 961, L37, \dodoi{10.3847/2041-8213/ad1b56}

\bibitem[{{Carleton} {et~al.}(2024{\natexlab{b}}){Carleton}, {Willner}, {Ellsworth-Bowers}, {Windhorst}, {Cohen}, {Conselice}, {Diego}, {Zitrin}, {Archer}, {McIntyre}, {Kamieneski}, {Jansen}, {Summers}, {D'Silva}, {Koekemoer}, {Coe}, {Driver}, {Frye}, {Grogin}, {Marshall}, {Nonino}, {Pirzkal}, {Robotham}, {Ryan}, {Ortiz}, {Tompkins}, {Willmer}, {Yan}, \& {Holwerda}}]{Carleton2024b}
{Carleton}, T., {Willner}, S.~P., {Ellsworth-Bowers}, T., {et~al.} 2024{\natexlab{b}}, Research Notes of the American Astronomical Society, 8, 181, \dodoi{10.3847/2515-5172/ad61d9}

\bibitem[{{Carlsten} {et~al.}(2019){Carlsten}, {Beaton}, {Greco}, \& {Greene}}]{Carlsten2019}
{Carlsten}, S.~G., {Beaton}, R.~L., {Greco}, J.~P., \& {Greene}, J.~E. 2019, \apj, 879, 13, \dodoi{10.3847/1538-4357/ab22c1}

\bibitem[{{Carlsten} {et~al.}(2022){Carlsten}, {Greene}, {Beaton}, {Danieli}, \& {Greco}}]{CarlstenELVES2022}
{Carlsten}, S.~G., {Greene}, J.~E., {Beaton}, R.~L., {Danieli}, S., \& {Greco}, J.~P. 2022, \apj, 933, 47, \dodoi{10.3847/1538-4357/ac6fd7}

\bibitem[{{Carlsten} {et~al.}(2021){Carlsten}, {Greene}, {Greco}, {Beaton}, \& {Kado-Fong}}]{ELVES-I}
{Carlsten}, S.~G., {Greene}, J.~E., {Greco}, J.~P., {Beaton}, R.~L., \& {Kado-Fong}, E. 2021, \apj, 922, 267, \dodoi{10.3847/1538-4357/ac2581}

\bibitem[{{Casey} {et~al.}(2023){Casey}, {Greco}, {Peter}, \& {Davis}}]{Casey2023}
{Casey}, K.~J., {Greco}, J.~P., {Peter}, A. H.~G., \& {Davis}, A.~B. 2023, \mnras, 520, 4715, \dodoi{10.1093/mnras/stad352}

\bibitem[{{Choi} {et~al.}(2016){Choi}, {Dotter}, {Conroy}, {Cantiello}, {Paxton}, \& {Johnson}}]{Choi2016}
{Choi}, J., {Dotter}, A., {Conroy}, C., {et~al.} 2016, \apj, 823, 102, \dodoi{10.3847/0004-637X/823/2/102}

\bibitem[{{Christensen} {et~al.}(2024){Christensen}, {Brooks}, {Munshi}, {Riggs}, {Van Nest}, {Akins}, {Quinn}, \& {Chamberland}}]{Christensen2024}
{Christensen}, C.~R., {Brooks}, A.~M., {Munshi}, F., {et~al.} 2024, \apj, 961, 236, \dodoi{10.3847/1538-4357/ad0c5a}

\bibitem[{{Cohen} {et~al.}(2018){Cohen}, {van Dokkum}, {Danieli}, {Romanowsky}, {Abraham}, {Merritt}, {Zhang}, {Mowla}, {Kruijssen}, {Conroy}, \& {Wasserman}}]{Cohen2018}
{Cohen}, Y., {van Dokkum}, P., {Danieli}, S., {et~al.} 2018, \apj, 868, 96, \dodoi{10.3847/1538-4357/aae7c8}

\bibitem[{{Crnojevi{\'c}} {et~al.}(2016){Crnojevi{\'c}}, {Sand}, {Zaritsky}, {Spekkens}, {Willman}, \& {Hargis}}]{Crnojevic2016}
{Crnojevi{\'c}}, D., {Sand}, D.~J., {Zaritsky}, D., {et~al.} 2016, \apjl, 824, L14, \dodoi{10.3847/2041-8205/824/1/L14}

\bibitem[{{Danieli} {et~al.}(2018){Danieli}, {van Dokkum}, \& {Conroy}}]{Danieli2018}
{Danieli}, S., {van Dokkum}, P., \& {Conroy}, C. 2018, \apj, 856, 69, \dodoi{10.3847/1538-4357/aaadfb}

\bibitem[{{Dey} {et~al.}(2019){Dey}, {Schlegel}, {Lang}, {Blum}, {Burleigh}, {Fan}, {Findlay}, {Finkbeiner}, {Herrera}, {Juneau}, {Landriau}, {Levi}, {McGreer}, {Meisner}, {Myers}, {Moustakas}, {Nugent}, {Patej}, {Schlafly}, {Walker}, {Valdes}, {Weaver}, {Y{\`e}che}, {Zou}, {Zhou}, {Abareshi}, {Abbott}, {Abolfathi}, {Aguilera}, {Alam}, {Allen}, {Alvarez}, {Annis}, {Ansarinejad}, {Aubert}, {Beechert}, {Bell}, {BenZvi}, {Beutler}, {Bielby}, {Bolton}, {Brice{\~n}o}, {Buckley-Geer}, {Butler}, {Calamida}, {Carlberg}, {Carter}, {Casas}, {Castander}, {Choi}, {Comparat}, {Cukanovaite}, {Delubac}, {DeVries}, {Dey}, {Dhungana}, {Dickinson}, {Ding}, {Donaldson}, {Duan}, {Duckworth}, {Eftekharzadeh}, {Eisenstein}, {Etourneau}, {Fagrelius}, {Farihi}, {Fitzpatrick}, {Font-Ribera}, {Fulmer}, {G{\"a}nsicke}, {Gaztanaga}, {George}, {Gerdes}, {Gontcho}, {Gorgoni}, {Green}, {Guy}, {Harmer}, {Hernandez}, {Honscheid}, {Huang}, {James}, {Jannuzi}, {Jiang}, {Joyce}, {Karcher}, {Karkar}, {Kehoe}, {Kneib}, {Kueter-Young}, {Lan},
  {Lauer}, {Le Guillou}, {Le Van Suu}, {Lee}, {Lesser}, {Perreault Levasseur}, {Li}, {Mann}, {Marshall}, {Mart{\'\i}nez-V{\'a}zquez}, {Martini}, {du Mas des Bourboux}, {McManus}, {Meier}, {M{\'e}nard}, {Metcalfe}, {Mu{\~n}oz-Guti{\'e}rrez}, {Najita}, {Napier}, {Narayan}, {Newman}, {Nie}, {Nord}, {Norman}, {Olsen}, {Paat}, {Palanque-Delabrouille}, {Peng}, {Poppett}, {Poremba}, {Prakash}, {Rabinowitz}, {Raichoor}, {Rezaie}, {Robertson}, {Roe}, {Ross}, {Ross}, {Rudnick}, {Safonova}, {Saha}, {S{\'a}nchez}, {Savary}, {Schweiker}, {Scott}, {Seo}, {Shan}, {Silva}, {Slepian}, {Soto}, {Sprayberry}, {Staten}, {Stillman}, {Stupak}, {Summers}, {Sien Tie}, {Tirado}, {Vargas-Maga{\~n}a}, {Vivas}, {Wechsler}, {Williams}, {Yang}, {Yang}, {Yapici}, {Zaritsky}, {Zenteno}, {Zhang}, {Zhang}, {Zhou}, \& {Zhou}}]{Dey2019}
{Dey}, A., {Schlegel}, D.~J., {Lang}, D., {et~al.} 2019, \aj, 157, 168, \dodoi{10.3847/1538-3881/ab089d}

\bibitem[{{Dotter}(2016)}]{Dotter2016}
{Dotter}, A. 2016, \apjs, 222, 8, \dodoi{10.3847/0067-0049/222/1/8}

\bibitem[{{Dressler} {et~al.}(2011){Dressler}, {Bigelow}, {Hare}, {Sutin}, {Thompson}, {Burley}, {Epps}, {Oemler}, {Bagish}, {Birk}, {Clardy}, {Gunnels}, {Kelson}, {Shectman}, \& {Osip}}]{Dressler2011}
{Dressler}, A., {Bigelow}, B., {Hare}, T., {et~al.} 2011, \pasp, 123, 288, \dodoi{10.1086/658908}

\bibitem[{{Drlica-Wagner} {et~al.}(2021){Drlica-Wagner}, {Carlin}, {Nidever}, {Ferguson}, {Kuropatkin}, {Adam{\'o}w}, {Cerny}, {Choi}, {Esteves}, {Mart{\'\i}nez-V{\'a}zquez}, {Mau}, {Miller}, {Mutlu-Pakdil}, {Neilsen}, {Olsen}, {Pace}, {Riley}, {Sakowska}, {Sand}, {Santana-Silva}, {Tollerud}, {Tucker}, {Vivas}, {Zaborowski}, {Zenteno}, {Abbott}, {Allam}, {Bechtol}, {Bell}, {Bell}, {Bilaji}, {Bom}, {Carballo-Bello}, {Crnojevi{\'c}}, {Cioni}, {Diaz-Ocampo}, {de Boer}, {Erkal}, {Gruendl}, {Hernandez-Lang}, {Hughes}, {James}, {Johnson}, {Li}, {Mao}, {Mart{\'\i}nez-Delgado}, {Massana}, {McNanna}, {Morgan}, {Nadler}, {No{\"e}l}, {Palmese}, {Peter}, {Rykoff}, {S{\'a}nchez}, {Shipp}, {Simon}, {Smercina}, {Soares-Santos}, {Stringfellow}, {Tavangar}, {van der Marel}, {Walker}, {Wechsler}, {Wu}, {Yanny}, {Fitzpatrick}, {Huang}, {Jacques}, {Nikutta}, {Scott}, \& {Astro Data Lab}}]{DELVE2021}
{Drlica-Wagner}, A., {Carlin}, J.~L., {Nidever}, D.~L., {et~al.} 2021, \apjs, 256, 2, \dodoi{10.3847/1538-4365/ac079d}

\bibitem[{{Drlica-Wagner} {et~al.}(2022){Drlica-Wagner}, {Ferguson}, {Adam{\'o}w}, {Aguena}, {Allam}, {Andrade-Oliveira}, {Bacon}, {Bechtol}, {Bell}, {Bertin}, {Bilaji}, {Bocquet}, {Bom}, {Brooks}, {Burke}, {Carballo-Bello}, {Carlin}, {Carnero Rosell}, {Carrasco Kind}, {Carretero}, {Castander}, {Cerny}, {Chang}, {Choi}, {Conselice}, {Costanzi}, {Crnojevi{\'c}}, {da Costa}, {de Vicente}, {Desai}, {Esteves}, {Everett}, {Ferrero}, {Fitzpatrick}, {Flaugher}, {Friedel}, {Frieman}, {Garc{\'\i}a-Bellido}, {Gatti}, {Gaztanaga}, {Gerdes}, {Gruen}, {Gruendl}, {Gschwend}, {Hartley}, {Hernandez-Lang}, {Hinton}, {Hollowood}, {Honscheid}, {Hughes}, {Jacques}, {James}, {Johnson}, {Kuehn}, {Kuropatkin}, {Lahav}, {Li}, {Lidman}, {Lin}, {March}, {Marshall}, {Mart{\'\i}nez-Delgado}, {Mart{\'\i}nez-V{\'a}zquez}, {Massana}, {Mau}, {McNanna}, {Melchior}, {Menanteau}, {Miller}, {Miquel}, {Mohr}, {Morgan}, {Mutlu-Pakdil}, {Mu{\~n}oz}, {Neilsen}, {Nidever}, {Nikutta}, {Nilo Castellon}, {No{\"e}l}, {Ogando}, {Olsen}, {Pace},
  {Palmese}, {Paz-Chinch{\'o}n}, {Pereira}, {Pieres}, {Plazas Malag{\'o}n}, {Prat}, {Riley}, {Rodriguez-Monroy}, {Romer}, {Roodman}, {Sako}, {Sakowska}, {Sanchez}, {S{\'a}nchez}, {Sand}, {Santana-Silva}, {Santiago}, {Schubnell}, {Serrano}, {Sevilla-Noarbe}, {Simon}, {Smith}, {Soares-Santos}, {Stringfellow}, {Suchyta}, {Suson}, {Tan}, {Tarle}, {Tavangar}, {Thomas}, {To}, {Tollerud}, {Troxel}, {Tucker}, {Varga}, {Vivas}, {Walker}, {Weller}, {Wilkinson}, {Wu}, {Yanny}, {Zaborowski}, {Zenteno}, {Delve Collaboration}, {Des Collaboration}, \& {Astro Data Lab}}]{DELVE2022}
{Drlica-Wagner}, A., {Ferguson}, P.~S., {Adam{\'o}w}, M., {et~al.} 2022, \apjs, 261, 38, \dodoi{10.3847/1538-4365/ac78eb}

\bibitem[{{Dumont} {et~al.}(2024){Dumont}, {Seth}, {Strader}, {Sand}, {Voggel}, {Hughes}, {Crnojevi{\'c}}, {Forbes}, {Mateo}, \& {Pearson}}]{Dumont2024}
{Dumont}, A., {Seth}, A.~C., {Strader}, J., {et~al.} 2024, \aap, 685, A132, \dodoi{10.1051/0004-6361/202347243}

\bibitem[{{Efstathiou}(1992)}]{Efstathiou1992}
{Efstathiou}, G. 1992, \mnras, 256, 43P, \dodoi{10.1093/mnras/256.1.43P}

\bibitem[{{El-Badry} {et~al.}(2018){El-Badry}, {Quataert}, {Wetzel}, {Hopkins}, {Weisz}, {Chan}, {Fitts}, {Boylan-Kolchin}, {Kere{\v{s}}}, {Faucher-Gigu{\`e}re}, \& {Garrison-Kimmel}}]{ElBadry2018}
{El-Badry}, K., {Quataert}, E., {Wetzel}, A., {et~al.} 2018, \mnras, 473, 1930, \dodoi{10.1093/mnras/stx2482}

\bibitem[{{Erwin}(2015)}]{imfit}
{Erwin}, P. 2015, \apj, 799, 226, \dodoi{10.1088/0004-637X/799/2/226}

\bibitem[{{Geha} {et~al.}(2012){Geha}, {Blanton}, {Yan}, \& {Tinker}}]{Geha2012}
{Geha}, M., {Blanton}, M.~R., {Yan}, R., \& {Tinker}, J.~L. 2012, \apj, 757, 85, \dodoi{10.1088/0004-637X/757/1/85}

\bibitem[{{Greco} \& {Danieli}(2022)}]{artpop}
{Greco}, J.~P., \& {Danieli}, S. 2022, \apj, 941, 26, \dodoi{10.3847/1538-4357/ac75b7}

\bibitem[{{Greco} {et~al.}(2018){Greco}, {Goulding}, {Greene}, {Strauss}, {Huang}, {Kim}, \& {Komiyama}}]{Greco2018b}
{Greco}, J.~P., {Goulding}, A.~D., {Greene}, J.~E., {et~al.} 2018, \apj, 866, 112, \dodoi{10.3847/1538-4357/aae0f4}

\bibitem[{{Greco} {et~al.}(2021){Greco}, {van Dokkum}, {Danieli}, {Carlsten}, \& {Conroy}}]{Greco2021}
{Greco}, J.~P., {van Dokkum}, P., {Danieli}, S., {Carlsten}, S.~G., \& {Conroy}, C. 2021, \apj, 908, 24, \dodoi{10.3847/1538-4357/abd030}

\bibitem[{{Greene} {et~al.}(2023){Greene}, {Danieli}, {Carlsten}, {Beaton}, {Jiang}, \& {Li}}]{Greene2023}
{Greene}, J.~E., {Danieli}, S., {Carlsten}, S., {et~al.} 2023, \apj, 949, 94, \dodoi{10.3847/1538-4357/acc58c}

\bibitem[{{Gunn} \& {Gott}(1972)}]{GunnGott1972}
{Gunn}, J.~E., \& {Gott}, J.~Richard, I. 1972, \apj, 176, 1, \dodoi{10.1086/151605}

\bibitem[{Harris {et~al.}(2020)Harris, Millman, van~der Walt, Gommers, Virtanen, Cournapeau, Wieser, Taylor, Berg, Smith, Kern, Picus, Hoyer, van Kerkwijk, Brett, Haldane, del R{'{\i}}o, Wiebe, Peterson, G{'{e}}rard-Marchant, Sheppard, Reddy, Weckesser, Abbasi, Gohlke, \& Oliphant}]{Numpy}
Harris, C.~R., Millman, K.~J., van~der Walt, S.~J., {et~al.} 2020, Nature, 585, 357, \dodoi{10.1038/s41586-020-2649-2}

\bibitem[{{Hopkins} {et~al.}(2012){Hopkins}, {Quataert}, \& {Murray}}]{Hopkins2012}
{Hopkins}, P.~F., {Quataert}, E., \& {Murray}, N. 2012, \mnras, 421, 3522, \dodoi{10.1111/j.1365-2966.2012.20593.x}

\bibitem[{{Huchtmeier} {et~al.}(2001){Huchtmeier}, {Karachentsev}, \& {Karachentseva}}]{Huchtmeier2001}
{Huchtmeier}, W.~K., {Karachentsev}, I.~D., \& {Karachentseva}, V.~E. 2001, \aap, 377, 801, \dodoi{10.1051/0004-6361:20011138}

\bibitem[{{Hunter}(2007)}]{matplotlib}
{Hunter}, J.~D. 2007, Computing in Science Engineering, 9, 90, \dodoi{10.1109/MCSE.2007.55}

\bibitem[{{Iglesias-P{\'a}ramo} {et~al.}(2006){Iglesias-P{\'a}ramo}, {Buat}, {Takeuchi}, {Xu}, {Boissier}, {Boselli}, {Burgarella}, {Madore}, {Gil de Paz}, {Bianchi}, {Barlow}, {Byun}, {Donas}, {Forster}, {Friedman}, {Heckman}, {Jelinski}, {Lee}, {Malina}, {Martin}, {Milliard}, {Morrissey}, {Neff}, {Rich}, {Schiminovich}, {Seibert}, {Siegmund}, {Small}, {Szalay}, {Welsh}, \& {Wyder}}]{Iglesias2006}
{Iglesias-P{\'a}ramo}, J., {Buat}, V., {Takeuchi}, T.~T., {et~al.} 2006, \apjs, 164, 38, \dodoi{10.1086/502628}

\bibitem[{{Into} \& {Portinari}(2013)}]{Into2013}
{Into}, T., \& {Portinari}, L. 2013, \mnras, 430, 2715, \dodoi{10.1093/mnras/stt071}

\bibitem[{{Ivezi{\'c}} {et~al.}(2019){Ivezi{\'c}}, {Kahn}, {Tyson}, {Abel}, {Acosta}, {Allsman}, {Alonso}, {AlSayyad}, {Anderson}, {Andrew}, {Angel}, {Angeli}, {Ansari}, {Antilogus}, {Araujo}, {Armstrong}, {Arndt}, {Astier}, {Aubourg}, {Auza}, {Axelrod}, {Bard}, {Barr}, {Barrau}, {Bartlett}, {Bauer}, {Bauman}, {Baumont}, {Bechtol}, {Bechtol}, {Becker}, {Becla}, {Beldica}, {Bellavia}, {Bianco}, {Biswas}, {Blanc}, {Blazek}, {Blandford}, {Bloom}, {Bogart}, {Bond}, {Booth}, {Borgland}, {Borne}, {Bosch}, {Boutigny}, {Brackett}, {Bradshaw}, {Brandt}, {Brown}, {Bullock}, {Burchat}, {Burke}, {Cagnoli}, {Calabrese}, {Callahan}, {Callen}, {Carlin}, {Carlson}, {Chandrasekharan}, {Charles-Emerson}, {Chesley}, {Cheu}, {Chiang}, {Chiang}, {Chirino}, {Chow}, {Ciardi}, {Claver}, {Cohen-Tanugi}, {Cockrum}, {Coles}, {Connolly}, {Cook}, {Cooray}, {Covey}, {Cribbs}, {Cui}, {Cutri}, {Daly}, {Daniel}, {Daruich}, {Daubard}, {Daues}, {Dawson}, {Delgado}, {Dellapenna}, {de Peyster}, {de Val-Borro}, {Digel}, {Doherty}, {Dubois},
  {Dubois-Felsmann}, {Durech}, {Economou}, {Eifler}, {Eracleous}, {Emmons}, {Fausti Neto}, {Ferguson}, {Figueroa}, {Fisher-Levine}, {Focke}, {Foss}, {Frank}, {Freemon}, {Gangler}, {Gawiser}, {Geary}, {Gee}, {Geha}, {Gessner}, {Gibson}, {Gilmore}, {Glanzman}, {Glick}, {Goldina}, {Goldstein}, {Goodenow}, {Graham}, {Gressler}, {Gris}, {Guy}, {Guyonnet}, {Haller}, {Harris}, {Hascall}, {Haupt}, {Hernandez}, {Herrmann}, {Hileman}, {Hoblitt}, {Hodgson}, {Hogan}, {Howard}, {Huang}, {Huffer}, {Ingraham}, {Innes}, {Jacoby}, {Jain}, {Jammes}, {Jee}, {Jenness}, {Jernigan}, {Jevremovi{\'c}}, {Johns}, {Johnson}, {Johnson}, {Jones}, {Juramy-Gilles}, {Juri{\'c}}, {Kalirai}, {Kallivayalil}, {Kalmbach}, {Kantor}, {Karst}, {Kasliwal}, {Kelly}, {Kessler}, {Kinnison}, {Kirkby}, {Knox}, {Kotov}, {Krabbendam}, {Krughoff}, {Kub{\'a}nek}, {Kuczewski}, {Kulkarni}, {Ku}, {Kurita}, {Lage}, {Lambert}, {Lange}, {Langton}, {Le Guillou}, {Levine}, {Liang}, {Lim}, {Lintott}, {Long}, {Lopez}, {Lotz}, {Lupton}, {Lust}, {MacArthur}, {Mahabal},
  {Mandelbaum}, {Markiewicz}, {Marsh}, {Marshall}, {Marshall}, {May}, {McKercher}, {McQueen}, {Meyers}, {Migliore}, {Miller}, {Mills}, {Miraval}, {Moeyens}, {Moolekamp}, {Monet}, {Moniez}, {Monkewitz}, {Montgomery}, {Morrison}, {Mueller}, {Muller}, {Mu{\~n}oz Arancibia}, {Neill}, {Newbry}, {Nief}, {Nomerotski}, {Nordby}, {O'Connor}, {Oliver}, {Olivier}, {Olsen}, {O'Mullane}, {Ortiz}, {Osier}, {Owen}, {Pain}, {Palecek}, {Parejko}, {Parsons}, {Pease}, {Peterson}, {Peterson}, {Petravick}, {Libby Petrick}, {Petry}, {Pierfederici}, {Pietrowicz}, {Pike}, {Pinto}, {Plante}, {Plate}, {Plutchak}, {Price}, {Prouza}, {Radeka}, {Rajagopal}, {Rasmussen}, {Regnault}, {Reil}, {Reiss}, {Reuter}, {Ridgway}, {Riot}, {Ritz}, {Robinson}, {Roby}, {Roodman}, {Rosing}, {Roucelle}, {Rumore}, {Russo}, {Saha}, {Sassolas}, {Schalk}, {Schellart}, {Schindler}, {Schmidt}, {Schneider}, {Schneider}, {Schoening}, {Schumacher}, {Schwamb}, {Sebag}, {Selvy}, {Sembroski}, {Seppala}, {Serio}, {Serrano}, {Shaw}, {Shipsey}, {Sick}, {Silvestri},
  {Slater}, {Smith}, {Smith}, {Sobhani}, {Soldahl}, {Storrie-Lombardi}, {Stover}, {Strauss}, {Street}, {Stubbs}, {Sullivan}, {Sweeney}, {Swinbank}, {Szalay}, {Takacs}, {Tether}, {Thaler}, {Thayer}, {Thomas}, {Thornton}, {Thukral}, {Tice}, {Trilling}, {Turri}, {Van Berg}, {Vanden Berk}, {Vetter}, {Virieux}, {Vucina}, {Wahl}, {Walkowicz}, {Walsh}, {Walter}, {Wang}, {Wang}, {Warner}, {Wiecha}, {Willman}, {Winters}, {Wittman}, {Wolff}, {Wood-Vasey}, {Wu}, {Xin}, {Yoachim}, \& {Zhan}}]{LSST2019}
{Ivezi{\'c}}, {\v{Z}}., {Kahn}, S.~M., {Tyson}, J.~A., {et~al.} 2019, \apj, 873, 111, \dodoi{10.3847/1538-4357/ab042c}

\bibitem[{{Jiang} {et~al.}(2019){Jiang}, {Dekel}, {Freundlich}, {Romanowsky}, {Dutton}, {Macci{\`o}}, \& {Di Cintio}}]{Jiang2019}
{Jiang}, F., {Dekel}, A., {Freundlich}, J., {et~al.} 2019, \mnras, 487, 5272, \dodoi{10.1093/mnras/stz1499}

\bibitem[{Jones {et~al.}(2001)Jones, Oliphant, Peterson, {et~al.}}]{scipy}
Jones, E., Oliphant, T., Peterson, P., {et~al.} 2001, {SciPy}: Open source scientific tools for {Python}.
\newblock \url{http://www.scipy.org/}

\bibitem[{{Kado-Fong} {et~al.}(2024){Kado-Fong}, {Robinson}, {Nyland}, {Greene}, {Suess}, {Stierwalt}, \& {Beaton}}]{Kado-Fong2024}
{Kado-Fong}, E., {Robinson}, A., {Nyland}, K., {et~al.} 2024, \apj, 963, 37, \dodoi{10.3847/1538-4357/ad18cb}

\bibitem[{{Kalberla} \& {Haud}(2015)}]{Kalberla2015}
{Kalberla}, P.~M.~W., \& {Haud}, U. 2015, \aap, 578, A78, \dodoi{10.1051/0004-6361/201525859}

\bibitem[{{Karachentsev} {et~al.}(2015){Karachentsev}, {Kniazev}, \& {Sharina}}]{Karachentsev2015_KKs3}
{Karachentsev}, I.~D., {Kniazev}, A.~Y., \& {Sharina}, M.~E. 2015, Astronomische Nachrichten, 336, 707, \dodoi{10.1002/asna.201512207}

\bibitem[{{Karachentsev} {et~al.}(2013){Karachentsev}, {Makarov}, \& {Kaisina}}]{Karachentsev2013}
{Karachentsev}, I.~D., {Makarov}, D.~I., \& {Kaisina}, E.~I. 2013, \aj, 145, 101, \dodoi{10.1088/0004-6256/145/4/101}

\bibitem[{{Karachentsev} {et~al.}(2018){Karachentsev}, {Makarova}, {Tully}, {Rizzi}, \& {Shaya}}]{Karachentsev2018}
{Karachentsev}, I.~D., {Makarova}, L.~N., {Tully}, R.~B., {Rizzi}, L., \& {Shaya}, E.~J. 2018, \apj, 858, 62, \dodoi{10.3847/1538-4357/aabaf1}

\bibitem[{{Karachentsev} {et~al.}(2007){Karachentsev}, {Tully}, {Dolphin}, {Sharina}, {Makarova}, {Makarov}, {Sakai}, {Shaya}, {Kashibadze}, {Karachentseva}, \& {Rizzi}}]{Karachentsev2007}
{Karachentsev}, I.~D., {Tully}, R.~B., {Dolphin}, A., {et~al.} 2007, \aj, 133, 504, \dodoi{10.1086/510125}

\bibitem[{{Karunakaran} {et~al.}(2022){Karunakaran}, {Spekkens}, {Carroll}, {Sand}, {Bennet}, {Crnojevi{\'c}}, {Jones}, \& {Mutlu-Pakd{\i}l}}]{Karunakaran2022}
{Karunakaran}, A., {Spekkens}, K., {Carroll}, R., {et~al.} 2022, \mnras, 516, 1741, \dodoi{10.1093/mnras/stac2329}

\bibitem[{{Karunakaran} {et~al.}(2021){Karunakaran}, {Spekkens}, {Oman}, {Simpson}, {Fattahi}, {Sand}, {Bennet}, {Crnojevi{\'c}}, {Frenk}, {G{\'o}mez}, {Grand}, {Jones}, {Marinacci}, {Mutlu-Pakdil}, {Navarro}, \& {Zaritsky}}]{Karunakaran2021}
{Karunakaran}, A., {Spekkens}, K., {Oman}, K.~A., {et~al.} 2021, \apjl, 916, L19, \dodoi{10.3847/2041-8213/ac0e3a}

\bibitem[{{Katz} {et~al.}(2020){Katz}, {Ramsoy}, {Rosdahl}, {Kimm}, {Blaizot}, {Haehnelt}, {Michel-Dansac}, {Garel}, {Laigle}, {Devriendt}, \& {Slyz}}]{Katz2020}
{Katz}, H., {Ramsoy}, M., {Rosdahl}, J., {et~al.} 2020, \mnras, 494, 2200, \dodoi{10.1093/mnras/staa639}

\bibitem[{{Kim} \& {Lee}(2021)}]{Kim2021}
{Kim}, Y.~J., \& {Lee}, M.~G. 2021, \apj, 923, 152, \dodoi{10.3847/1538-4357/ac2d94}

\bibitem[{{Kirby} {et~al.}(2013){Kirby}, {Cohen}, {Guhathakurta}, {Cheng}, {Bullock}, \& {Gallazzi}}]{Kirby2013}
{Kirby}, E.~N., {Cohen}, J.~G., {Guhathakurta}, P., {et~al.} 2013, \apj, 779, 102, \dodoi{10.1088/0004-637X/779/2/102}

\bibitem[{{Koribalski} {et~al.}(2020){Koribalski}, {Staveley-Smith}, {Westmeier}, {Serra}, {Spekkens}, {Wong}, {Lee-Waddell}, {Lagos}, {Obreschkow}, {Ryan-Weber}, {Zwaan}, {Kilborn}, {Bekiaris}, {Bekki}, {Bigiel}, {Boselli}, {Bosma}, {Catinella}, {Chauhan}, {Cluver}, {Colless}, {Courtois}, {Crain}, {de Blok}, {D{\'e}nes}, {Duffy}, {Elagali}, {Fluke}, {For}, {Heald}, {Henning}, {Hess}, {Holwerda}, {Howlett}, {Jarrett}, {Jones}, {Jones}, {J{\'o}zsa}, {Jurek}, {J{\"u}tte}, {Kamphuis}, {Karachentsev}, {Kerp}, {Kleiner}, {Kraan-Korteweg}, {L{\'o}pez-S{\'a}nchez}, {Madrid}, {Meyer}, {Mould}, {Murugeshan}, {Norris}, {Oh}, {Oosterloo}, {Popping}, {Putman}, {Reynolds}, {Rhee}, {Robotham}, {Ryder}, {Schr{\"o}der}, {Shao}, {Stevens}, {Taylor}, {van{\^A} der Hulst}, {Verdes-Montenegro}, {Wakker}, {Wang}, {Whiting}, {Winkel}, \& {Wolf}}]{WALLABY2020}
{Koribalski}, B.~S., {Staveley-Smith}, L., {Westmeier}, T., {et~al.} 2020, \apss, 365, 118, \dodoi{10.1007/s10509-020-03831-4}

\bibitem[{{Koss} {et~al.}(2022){Koss}, {Ricci}, {Trakhtenbrot}, {Oh}, {den Brok}, {Mej{\'\i}a-Restrepo}, {Stern}, {Privon}, {Treister}, {Powell}, {Mushotzky}, {Bauer}, {Ananna}, {Balokovi{\'c}}, {B{\"a}r}, {Becker}, {Bessiere}, {Burtscher}, {Caglar}, {Congiu}, {Evans}, {Harrison}, {Heida}, {Ichikawa}, {Kamraj}, {Lamperti}, {Pacucci}, {Ricci}, {Riffel}, {Rojas}, {Schawinski}, {Temple}, {Urry}, {Veilleux}, \& {Williams}}]{Koss2022}
{Koss}, M.~J., {Ricci}, C., {Trakhtenbrot}, B., {et~al.} 2022, \apjs, 261, 2, \dodoi{10.3847/1538-4365/ac6c05}

\bibitem[{{Kravtsov}(2013)}]{Kravtsov2013}
{Kravtsov}, A.~V. 2013, \apjl, 764, L31, \dodoi{10.1088/2041-8205/764/2/L31}

\bibitem[{{Kroupa}(2001)}]{Kroupa2001}
{Kroupa}, P. 2001, \mnras, 322, 231, \dodoi{10.1046/j.1365-8711.2001.04022.x}

\bibitem[{{Lang} {et~al.}(2010){Lang}, {Hogg}, {Mierle}, {Blanton}, \& {Roweis}}]{astrometry_net}
{Lang}, D., {Hogg}, D.~W., {Mierle}, K., {Blanton}, M., \& {Roweis}, S. 2010, \aj, 139, 1782, \dodoi{10.1088/0004-6256/139/5/1782}

\bibitem[{{Lee} {et~al.}(2011){Lee}, {Gil de Paz}, {Kennicutt}, {Bothwell}, {Dalcanton}, {Jos{\'e} G. Funes S.}, {Johnson}, {Sakai}, {Skillman}, {Tremonti}, \& {van Zee}}]{Lee2011}
{Lee}, J.~C., {Gil de Paz}, A., {Kennicutt}, Robert~C., J., {et~al.} 2011, \apjs, 192, 6, \dodoi{10.1088/0067-0049/192/1/6}

\bibitem[{{Li} {et~al.}(2023){Li}, {Greene}, {Greco}, {Huang}, {Melchior}, {Beaton}, {Casey}, {Danieli}, {Goulding}, {Joseph}, {Kado-Fong}, {Kim}, \& {MacArthur}}]{Li2023a}
{Li}, J., {Greene}, J.~E., {Greco}, J.~P., {et~al.} 2023, \apj, 955, 1, \dodoi{10.3847/1538-4357/ace829}

\bibitem[{{Makarov} {et~al.}(2012){Makarov}, {Makarova}, {Sharina}, {Uklein}, {Tikhonov}, {Guhathakurta}, {Kirby}, \& {Terekhova}}]{Makarov2012}
{Makarov}, D., {Makarova}, L., {Sharina}, M., {et~al.} 2012, \mnras, 425, 709, \dodoi{10.1111/j.1365-2966.2012.21581.x}

\bibitem[{{Manwadkar} \& {Kravtsov}(2022)}]{Manwadkar2022}
{Manwadkar}, V., \& {Kravtsov}, A.~V. 2022, \mnras, 516, 3944, \dodoi{10.1093/mnras/stac2452}

\bibitem[{{Martin} {et~al.}(2005){Martin}, {Fanson}, {Schiminovich}, {Morrissey}, {Friedman}, {Barlow}, {Conrow}, {Grange}, {Jelinsky}, {Milliard}, {Siegmund}, {Bianchi}, {Byun}, {Donas}, {Forster}, {Heckman}, {Lee}, {Madore}, {Malina}, {Neff}, {Rich}, {Small}, {Surber}, {Szalay}, {Welsh}, \& {Wyder}}]{Martin2005}
{Martin}, D.~C., {Fanson}, J., {Schiminovich}, D., {et~al.} 2005, \apjl, 619, L1, \dodoi{10.1086/426387}

\bibitem[{{McClure-Griffiths} {et~al.}(2009){McClure-Griffiths}, {Pisano}, {Calabretta}, {Ford}, {Lockman}, {Staveley-Smith}, {Kalberla}, {Bailin}, {Dedes}, {Janowiecki}, {Gibson}, {Murphy}, {Nakanishi}, \& {Newton-McGee}}]{McClure-Griffiths2009}
{McClure-Griffiths}, N.~M., {Pisano}, D.~J., {Calabretta}, M.~R., {et~al.} 2009, \apjs, 181, 398, \dodoi{10.1088/0067-0049/181/2/398}

\bibitem[{{McConnachie}(2012)}]{McConnachie2012}
{McConnachie}, A.~W. 2012, \aj, 144, 4, \dodoi{10.1088/0004-6256/144/1/4}

\bibitem[{{McConnachie} \& {Irwin}(2006)}]{McConnachie2006}
{McConnachie}, A.~W., \& {Irwin}, M.~J. 2006, \mnras, 365, 1263, \dodoi{10.1111/j.1365-2966.2005.09806.x}

\bibitem[{{McConnachie} {et~al.}(2008){McConnachie}, {Huxor}, {Martin}, {Irwin}, {Chapman}, {Fahlman}, {Ferguson}, {Ibata}, {Lewis}, {Richer}, \& {Tanvir}}]{McConnachie2008}
{McConnachie}, A.~W., {Huxor}, A., {Martin}, N.~F., {et~al.} 2008, \apj, 688, 1009, \dodoi{10.1086/591313}

\bibitem[{{McQuinn} {et~al.}(2024){McQuinn}, {B. Newman}, {Savino}, {Dolphin}, {Weisz}, {Williams}, {Boyer}, {Cohen}, {Correnti}, {Cole}, {Geha}, {Gennaro}, {Kallivayalil}, {Sandstrom}, {Skillman}, {Anderson}, {Bolatto}, {Boylan-Kolchin}, {Garling}, {Gilbert}, {Girardi}, {Kalirai}, {Mazzi}, {Pastorelli}, {Richstein}, \& {Warfield}}]{McQuinn2024}
{McQuinn}, K. B.~W., {B. Newman}, M.~J., {Savino}, A., {et~al.} 2024, \apj, 961, 16, \dodoi{10.3847/1538-4357/ad1105}

\bibitem[{{Mei} {et~al.}(2005){Mei}, {Blakeslee}, {Tonry}, {Jord{\'a}n}, {Peng}, {C{\^o}t{\'e}}, {Ferrarese}, {Merritt}, {Milosavljevi{\'c}}, \& {West}}]{Mei2005}
{Mei}, S., {Blakeslee}, J.~P., {Tonry}, J.~L., {et~al.} 2005, \apjs, 156, 113, \dodoi{10.1086/426544}

\bibitem[{{More} {et~al.}(2015){More}, {Diemer}, \& {Kravtsov}}]{More2015}
{More}, S., {Diemer}, B., \& {Kravtsov}, A.~V. 2015, \apj, 810, 36, \dodoi{10.1088/0004-637X/810/1/36}

\bibitem[{{Morrissey} {et~al.}(2007){Morrissey}, {Conrow}, {Barlow}, {Small}, {Seibert}, {Wyder}, {Budav{\'a}ri}, {Arnouts}, {Friedman}, {Forster}, {Martin}, {Neff}, {Schiminovich}, {Bianchi}, {Donas}, {Heckman}, {Lee}, {Madore}, {Milliard}, {Rich}, {Szalay}, {Welsh}, \& {Yi}}]{Morrissey2007}
{Morrissey}, P., {Conrow}, T., {Barlow}, T.~A., {et~al.} 2007, \apjs, 173, 682, \dodoi{10.1086/520512}

\bibitem[{{M{\"u}ller} {et~al.}(2017){M{\"u}ller}, {Jerjen}, \& {Binggeli}}]{Muller2017}
{M{\"u}ller}, O., {Jerjen}, H., \& {Binggeli}, B. 2017, \aap, 597, A7, \dodoi{10.1051/0004-6361/201628921}

\bibitem[{{M{\"u}ller} {et~al.}(2022){M{\"u}ller}, {Lelli}, {Famaey}, {Pawlowski}, {Fahrion}, {Rejkuba}, {Hilker}, \& {Jerjen}}]{Muller2022}
{M{\"u}ller}, O., {Lelli}, F., {Famaey}, B., {et~al.} 2022, \aap, 662, A57, \dodoi{10.1051/0004-6361/202142351}

\bibitem[{{M{\"u}ller} {et~al.}(2019){M{\"u}ller}, {Rejkuba}, {Pawlowski}, {Ibata}, {Lelli}, {Hilker}, \& {Jerjen}}]{Muller2019}
{M{\"u}ller}, O., {Rejkuba}, M., {Pawlowski}, M.~S., {et~al.} 2019, \aap, 629, A18, \dodoi{10.1051/0004-6361/201935807}

\bibitem[{{Okamoto} {et~al.}(2008){Okamoto}, {Gao}, \& {Theuns}}]{Okamoto2008}
{Okamoto}, T., {Gao}, L., \& {Theuns}, T. 2008, \mnras, 390, 920, \dodoi{10.1111/j.1365-2966.2008.13830.x}

\bibitem[{{Oke} \& {Gunn}(1983)}]{Oke1983}
{Oke}, J.~B., \& {Gunn}, J.~E. 1983, \apj, 266, 713, \dodoi{10.1086/160817}

\bibitem[{{Patel} {et~al.}(2018){Patel}, {Besla}, {Mandel}, \& {Sohn}}]{Patel2018}
{Patel}, E., {Besla}, G., {Mandel}, K., \& {Sohn}, S.~T. 2018, \apj, 857, 78, \dodoi{10.3847/1538-4357/aab78f}

\bibitem[{{Polzin} {et~al.}(2021){Polzin}, {van Dokkum}, {Danieli}, {Greco}, \& {Romanowsky}}]{Polzin2021}
{Polzin}, A., {van Dokkum}, P., {Danieli}, S., {Greco}, J.~P., \& {Romanowsky}, A.~J. 2021, \apjl, 914, L23, \dodoi{10.3847/2041-8213/ac024f}

\bibitem[{{Putman} {et~al.}(2021){Putman}, {Zheng}, {Price-Whelan}, {Grcevich}, {Johnson}, {Tollerud}, \& {Peek}}]{Putman2021}
{Putman}, M.~E., {Zheng}, Y., {Price-Whelan}, A.~M., {et~al.} 2021, \apj, 913, 53, \dodoi{10.3847/1538-4357/abe391}

\bibitem[{{Rees}(1986)}]{Rees1986}
{Rees}, M.~J. 1986, \mnras, 218, 25P, \dodoi{10.1093/mnras/218.1.25P}

\bibitem[{{Rey} {et~al.}(2020){Rey}, {Pontzen}, {Agertz}, {Orkney}, {Read}, \& {Rosdahl}}]{Rey2020}
{Rey}, M.~P., {Pontzen}, A., {Agertz}, O., {et~al.} 2020, \mnras, 497, 1508, \dodoi{10.1093/mnras/staa1640}

\bibitem[{{Ricotti} \& {Gnedin}(2005)}]{Ricotti2005}
{Ricotti}, M., \& {Gnedin}, N.~Y. 2005, \apj, 629, 259, \dodoi{10.1086/431415}

\bibitem[{{Rodr{\'\i}guez-Puebla} {et~al.}(2017){Rodr{\'\i}guez-Puebla}, {Primack}, {Avila-Reese}, \& {Faber}}]{Rodriguez-Puebla2017}
{Rodr{\'\i}guez-Puebla}, A., {Primack}, J.~R., {Avila-Reese}, V., \& {Faber}, S.~M. 2017, \mnras, 470, 651, \dodoi{10.1093/mnras/stx1172}

\bibitem[{{Samuel} {et~al.}(2022){Samuel}, {Wetzel}, {Santistevan}, {Tollerud}, {Moreno}, {Boylan-Kolchin}, {Bailin}, \& {Pardasani}}]{Samuel2022}
{Samuel}, J., {Wetzel}, A., {Santistevan}, I., {et~al.} 2022, \mnras, 514, 5276, \dodoi{10.1093/mnras/stac1706}

\bibitem[{{Sand} {et~al.}(2022){Sand}, {Mutlu-Pakdil}, {Jones}, {Karunakaran}, {Wang}, {Yang}, {Chiti}, {Bennet}, {Crnojevi{\'c}}, \& {Spekkens}}]{Sand2022}
{Sand}, D.~J., {Mutlu-Pakdil}, B., {Jones}, M.~G., {et~al.} 2022, \apjl, 935, L17, \dodoi{10.3847/2041-8213/ac85ee}

\bibitem[{{Santos-Santos} {et~al.}(2023){Santos-Santos}, {Navarro}, \& {McConnachie}}]{SantosSantos2023}
{Santos-Santos}, I. M.~E., {Navarro}, J.~F., \& {McConnachie}, A. 2023, \mnras, 520, 55, \dodoi{10.1093/mnras/stad085}

\bibitem[{{Saviane} {et~al.}(1996){Saviane}, {Held}, \& {Piotto}}]{Saviane1996}
{Saviane}, I., {Held}, E.~V., \& {Piotto}, G. 1996, \aap, 315, 40, \dodoi{10.48550/arXiv.astro-ph/9601165}

\bibitem[{{Schlafly} \& {Finkbeiner}(2011)}]{Schlafly2011}
{Schlafly}, E.~F., \& {Finkbeiner}, D.~P. 2011, \apj, 737, 103, \dodoi{10.1088/0004-637X/737/2/103}

\bibitem[{{Schlegel} {et~al.}(1998){Schlegel}, {Finkbeiner}, \& {Davis}}]{SFD1998}
{Schlegel}, D.~J., {Finkbeiner}, D.~P., \& {Davis}, M. 1998, \apj, 500, 525, \dodoi{10.1086/305772}

\bibitem[{{Shapiro} {et~al.}(2004){Shapiro}, {Iliev}, \& {Raga}}]{Shapiro2004}
{Shapiro}, P.~R., {Iliev}, I.~T., \& {Raga}, A.~C. 2004, \mnras, 348, 753, \dodoi{10.1111/j.1365-2966.2004.07364.x}

\bibitem[{{Simpson} {et~al.}(2018){Simpson}, {Grand}, {G{\'o}mez}, {Marinacci}, {Pakmor}, {Springel}, {Campbell}, \& {Frenk}}]{Simpson2018}
{Simpson}, C.~M., {Grand}, R. J.~J., {G{\'o}mez}, F.~A., {et~al.} 2018, \mnras, 478, 548, \dodoi{10.1093/mnras/sty774}

\bibitem[{{Slater} \& {Bell}(2014)}]{Slater2014}
{Slater}, C.~T., \& {Bell}, E.~F. 2014, \apj, 792, 141, \dodoi{10.1088/0004-637X/792/2/141}

\bibitem[{{Teyssier} {et~al.}(2012){Teyssier}, {Johnston}, \& {Kuhlen}}]{Teyssier2012}
{Teyssier}, M., {Johnston}, K.~V., \& {Kuhlen}, M. 2012, \mnras, 426, 1808, \dodoi{10.1111/j.1365-2966.2012.21793.x}

\bibitem[{{Tonry} \& {Schneider}(1988)}]{Tonry1988}
{Tonry}, J., \& {Schneider}, D.~P. 1988, \aj, 96, 807, \dodoi{10.1086/114847}

\bibitem[{{Tully} {et~al.}(2015){Tully}, {Libeskind}, {Karachentsev}, {Karachentseva}, {Rizzi}, \& {Shaya}}]{Tully2015}
{Tully}, R.~B., {Libeskind}, N.~I., {Karachentsev}, I.~D., {et~al.} 2015, \apjl, 802, L25, \dodoi{10.1088/2041-8205/802/2/L25}

\bibitem[{{Wang} {et~al.}(2020){Wang}, {Hammer}, {Rejkuba}, {Crnojevi{\'c}}, \& {Yang}}]{Wang2020}
{Wang}, J., {Hammer}, F., {Rejkuba}, M., {Crnojevi{\'c}}, D., \& {Yang}, Y. 2020, \mnras, 498, 2766, \dodoi{10.1093/mnras/staa2508}

\bibitem[{{Wang} {et~al.}(2024){Wang}, {Nadler}, {Mao}, {Wechsler}, {Abel}, {Behroozi}, {Geha}, {Asali}, {de los Reyes}, {Kado-Fong}, {Kallivayalil}, {Tollerud}, {Weiner}, \& {Wu}}]{Wang2024}
{Wang}, Y., {Nadler}, E.~O., {Mao}, Y.-Y., {et~al.} 2024, arXiv e-prints, arXiv:2404.14500, \dodoi{10.48550/arXiv.2404.14500}

\bibitem[{{Weisz} {et~al.}(2014){Weisz}, {Dolphin}, {Skillman}, {Holtzman}, {Gilbert}, {Dalcanton}, \& {Williams}}]{Weisz2014}
{Weisz}, D.~R., {Dolphin}, A.~E., {Skillman}, E.~D., {et~al.} 2014, \apj, 789, 147, \dodoi{10.1088/0004-637X/789/2/147}

\bibitem[{{Weisz} {et~al.}(2011){Weisz}, {Dalcanton}, {Williams}, {Gilbert}, {Skillman}, {Seth}, {Dolphin}, {McQuinn}, {Gogarten}, {Holtzman}, {Rosema}, {Cole}, {Karachentsev}, \& {Zaritsky}}]{Weisz2011}
{Weisz}, D.~R., {Dalcanton}, J.~J., {Williams}, B.~F., {et~al.} 2011, \apj, 739, 5, \dodoi{10.1088/0004-637X/739/1/5}

\bibitem[{{Wetzel} {et~al.}(2014){Wetzel}, {Tinker}, {Conroy}, \& {van den Bosch}}]{Wetzel2014}
{Wetzel}, A.~R., {Tinker}, J.~L., {Conroy}, C., \& {van den Bosch}, F.~C. 2014, \mnras, 439, 2687, \dodoi{10.1093/mnras/stu122}

\bibitem[{{Wetzel} {et~al.}(2015){Wetzel}, {Tollerud}, \& {Weisz}}]{Wetzel2015}
{Wetzel}, A.~R., {Tollerud}, E.~J., \& {Weisz}, D.~R. 2015, \apjl, 808, L27, \dodoi{10.1088/2041-8205/808/1/L27}

\bibitem[{{Willmer}(2018)}]{Willmer2018}
{Willmer}, C. N.~A. 2018, \apjs, 236, 47, \dodoi{10.3847/1538-4365/aabfdf}

\bibitem[{{Wyder} {et~al.}(2007){Wyder}, {Martin}, {Schiminovich}, {Seibert}, {Budav{\'a}ri}, {Treyer}, {Barlow}, {Forster}, {Friedman}, {Morrissey}, {Neff}, {Small}, {Bianchi}, {Donas}, {Heckman}, {Lee}, {Madore}, {Milliard}, {Rich}, {Szalay}, {Welsh}, \& {Yi}}]{Wyder2007}
{Wyder}, T.~K., {Martin}, D.~C., {Schiminovich}, D., {et~al.} 2007, \apjs, 173, 293, \dodoi{10.1086/521402}

\bibitem[{{Xi} {et~al.}(2018){Xi}, {Taylor}, {Massey}, {Rhodes}, {Koekemoer}, \& {Salvato}}]{Xi2018}
{Xi}, C., {Taylor}, J.~E., {Massey}, R.~J., {et~al.} 2018, \mnras, 478, 5336, \dodoi{10.1093/mnras/sty1333}

\end{thebibliography}
\bibliographystyle{aasjournal}

\newpage
\appendix 

\section{Measurement uncertainties and biases}\label{ap:bias}

The structural parameters of low surface brightness galaxies are typically obtained by fitting a \sersic model to the image. However, such measurements are subject to potential biases and uncertainties because of the low surface brightness nature of the target galaxies. The measurement could also be biased for those resolved and semi-resolved galaxies. Similar to \citet{Casey2023} and \citet{Li2023a}, we characterize the measurement biases and uncertainties by generating mock \sersic galaxy using \artpop \citep{artpop} and compare the measured properties with the ground truth. We use \artpop models rather than a smooth \sersic profile to fully mimic the SBF of Hedgehog and investigate how that would affect the measurement bias and uncertainty. 

Assuming the mock galaxies have similar properties (distance, total magnitude, effective radius, \sersic index, ellipticity) as Hedgehog and have an SSP with an age of 6 Gyr and metallicity of $\rm [Fe/H]=-1.75$, we generated 100 different realizations of the mock galaxies with random orientations. These mock galaxies were randomly injected into the Legacy Surveys DR10 coadded images within $5\arcmin$ from Hedgehog (i.e., within the same brick). We ran the \sersic fitting pipeline for each of these mock galaxies in a similar way to Hedgehog. 

The measurement bias is defined as $\Delta X = X_{\rm truth} - X_{\rm meas}$, and the measurement uncertainty $\sigma_X$ is the standard deviation of the measurement difference. The median biases and the uncertainties are listed in Table \ref{tab:bias_err}. The bias values are typically within the uncertainty range, except for the $g$-band magnitude $m_g$ and effective radius $r_{\rm eff}$. The derived biases and uncertainties are comparable to the values in \citet{ELVES-I} and \citet{Casey2023}. We then apply bias corrections to the measurement of Hedgehog. The properties of Hedgehog in Table \ref{tab:property} have already been corrected for measurement biases. 

However, the above procedure does not properly model the uncertainty introduced in sky background subtraction and image coaddition. As a consequence, the uncertainties reported here are likely underestimated. Injecting mock galaxies into raw data before sky subtraction would require a nontrivial amount of data reduction using the Legacy Surveys reduction pipeline, which is beyond the scope of this work.

\begin{table}
\centering
\label{tab:bias_err}
\caption{Measurement Biases and Uncertainties of the Structural Parameters. }
\begin{tabular}{lcc}
\toprule
Parameter & Uncertainty & Bias \\
\midrule
$m_g$ (mag) &  $\pm 0.06$ & $0.12$ \\
$g-r$ (mag) & $\pm 0.05$ & $-0.02$\\
$g-i$ (mag) & $\pm 0.05$ & $-0.03$\\
$r_{\rm eff}$ (arcsec) & $\pm 0.8$ & $-1.0$\\
$n_{\rm S\acute{e}rsic}$ &  $\pm 0.12$ & $-0.05$\\
Ellipticity & $\pm 0.03$ & $-0.01$ \\
\bottomrule
\end{tabular}
\end{table}

\begin{figure}
    \centering
    \includegraphics[width=1\linewidth]{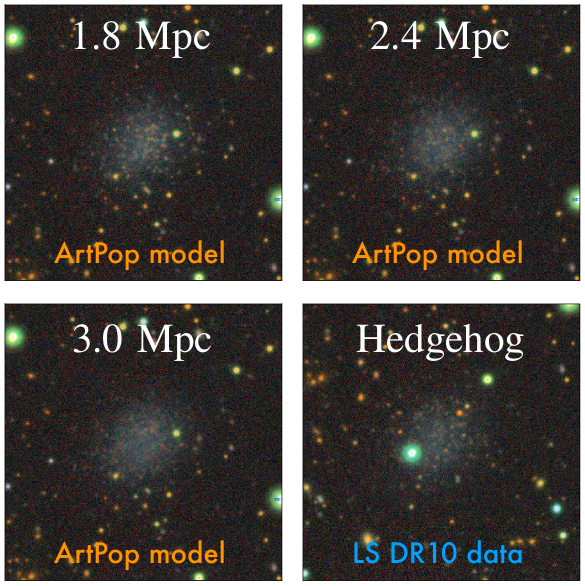}
    \caption{\code{ArtPop} models of Hedgehog at different distances. We generate these mock galaxies assuming an SSP with an age of 6 Gyr and metallicity of $\feh=-1.75$ and place them at different distances but fix the shape and surface brightness. The mock galaxy at 2.4 Mpc visually reproduces both the SBF signal and the integrated color of Hedgehog very well. The two galaxies at 1.8 Mpc and 3.0 Mpc show too strong and too weak SBF signals. 
    }
    \label{fig:artpop}
\end{figure}

\section{\textsc{ArtPop} models}\label{ap:artpop}
To further demonstrate that the measured distance in \S\ref{sec:distance} and the inferred stellar population in \S\ref{sec:sp} are reasonable, we generate mock \sersic galaxy images using an SSP with an age of 6 Gyr and metallicity of $\feh=-1.75$ using \code{ArtPop} \citep{artpop}. We place the mock galaxy at 1.8~Mpc, 2.4~Mpc, and 3.0~Mpc, fixing the surface brightness so that the mock galaxy at 2.4~Mpc has the same structural parameters as Hedgehog. The mock galaxy is then injected into the Legacy Surveys DR10 $griz$ data at $\alpha=\rm13h22m46.99s$, $\delta=\rm-20d52m30.36s$ ($1.5\arcmin$ from Hedgehog), where the data quality is very similar to the location of Hedgehog. The mock images are shown in Figure \ref{fig:artpop} together with Hedgehog. The mock galaxy at 2.4~Mpc is most similar to Hedgehog in terms of the SBF signal. The integrated color of the mock galaxies also matches the color of Hedgehog quite well.

\end{CJK*}
\end{document}